\ifdefined\pdftexversion\pdfoutput=1\fi  
\documentclass[12pt]{article}

\usepackage[margin=1in]{geometry}
\usepackage[T1]{fontenc}
\usepackage{mathptmx}
\usepackage{booktabs}
\usepackage{tabularray}
\UseTblrLibrary{booktabs}
\usepackage[listings,skins]{tcolorbox}
\usepackage{amsmath}
\usepackage{url}
\usepackage{listings}
\usepackage{tikz}
\usetikzlibrary{arrows.meta,positioning,fit,backgrounds}

\tikzset{
  aircraft/.pic={
    \fill (0,0.24)
       -- (0.03,0.14) -- (0.035,0.03)
       -- (0.26,-0.10) -- (0.26,-0.145)
       -- (0.035,-0.075) -- (0.035,-0.175)
       -- (0.12,-0.245) -- (0.12,-0.275)
       -- (0,-0.225)
       -- (-0.12,-0.275) -- (-0.12,-0.245)
       -- (-0.035,-0.175) -- (-0.035,-0.075)
       -- (-0.26,-0.145) -- (-0.26,-0.10)
       -- (-0.035,0.03) -- (-0.03,0.14) -- cycle;
  }
}
\usepackage{pifont}
\DeclareRobustCommand{\chnum}[1]{\ding{\ifcase#1\or172\or173\or174\or175\fi}}

\usepackage[hidelinks]{hyperref}
\usepackage{titling}

\predate{}
\postdate{}
\usepackage[htt]{hyphenat}

\lstdefinestyle{paper}{
  basicstyle=\ttfamily\scriptsize,
  columns=fullflexible,
  keepspaces=true,
  breaklines=true,
  commentstyle=\itshape\color{black!55},
  morecomment=[l]{//},
  frame=none,
  xleftmargin=0pt,
  aboveskip=0pt, belowskip=0pt,
}
\tcbset{
  codebox/.style={
    listing only, listing style=paper,
    boxrule=0.4pt, arc=2.5pt, colback=black!2, colframe=black!30,
    left=5pt, right=5pt, top=4pt, bottom=4pt, boxsep=0pt,
  },
}

\SetTblrInner{rowsep=1.6pt, colsep=5pt}
\NewTblrTheme{paper}{\SetTblrStyle{firsthead}{font=\bfseries}}

\title{Oracles That Cannot Fail:\\
Anchoring and the Expectation That Moves With the Fault}

\author{Arquimedes Canedo\\
\texttt{canedo@sector.mx}\\
\small ORCID \href{https://orcid.org/0000-0003-3506-6563}{0000-0003-3506-6563}}

\date{}

\begin{document}
\maketitle
\vspace{-2em}

\begin{abstract}
A test oracle that obtains its expected value from the system it is judging cannot
fail. If a fault moves measurement and expectation together the comparison cancels
exactly, and no generated input will reveal it. The defect is in the oracle and
not in the input space. We call this \emph{oracle anchoring}. An expectation is specification-anchored when composed from values
fixed outside the code under mutation, and state-anchored when it flows, directly
or transitively, from that code. The expected-value form is named in the test-smell
literature but not measured in any study we retrieved. We name three further channels by which such a
value reaches a verdict, restrict the predicate to values flowing from the mutate
target, and measure it. The subject is a deployed air traffic control simulator with 12 model-free
property suites. Across 4 modules and 366
mutants these add 3 mutants of detection over the hand-written tests, while remaining 6 to 33 times as efficient per test. We then intervene
three times, predicting each outcome first. Re-anchoring one holding oracle on published
procedure, changing no production code, recovers 8 of 46; state-anchoring a healthy debounce oracle costs 4 of 19;
and a reference model on that population kills exactly what specification
anchoring kills, placing the risk in anchoring and not in model-freedom. Of 6 instances
ablated, the two sizing their comparison carry 11 of the 12 recovered mutants. The
published smell rule would revert our repair. Writing the oracle this analysis said was
missing then exposed two defects deployment had not surfaced. All measurements come from one system by one author.
\end{abstract}

\vspace{0.5em}

\section{Introduction}
\label{sec:intro}

An air traffic control simulator is judged on whether its aircraft behave
plausibly, and the behaviors that carry that judgment are almost all temporal. An
eligibility prompt must persist across a brief geometric dropout but not
indefinitely. An approach must progress through its phases in order and without
skipping. A holding pattern must circulate over its fix rather than beside it. A
resolution advisory must override every controller command for as long as it is
active and no longer. These are not properties of a single output value. They are
properties of a trajectory through state, and they fail in orderings, in
time-step chunkings, and in starting geometries that a software engineer does not think to
enumerate.

\begin{figure}[t]
\centering
\begin{tikzpicture}[
  font=\footnotesize,
  trk/.style={thick},
  dim/.style={{Stealth[length=1.6mm]}-{Stealth[length=1.6mm]}},
  crs/.style={dashed},
  fx/.style={circle, fill, inner sep=1.5pt},
]
\def\g{0.08}
\begin{scope}
  \def\r{0.75}\def\L{2.8}
  \draw[crs] (-\L-\r-0.55,0) -- (\r+0.55,0);
  \node[anchor=east, font=\scriptsize] at (-\L-\r-0.6,0) {course};
  \draw[trk,-{Stealth[length=2mm]}] (-\L,0) -- (-\L/2,0);
  \draw[trk] (-\L/2,0) -- (0,0);
  \draw[trk] (0,0) arc[start angle=90, end angle=-90, radius=\r];
  \draw[trk,-{Stealth[length=2mm]}] (0,-2*\r) -- (-\L/2,-2*\r);
  \draw[trk] (-\L/2,-2*\r) -- (-\L,-2*\r);
  \draw[trk] (-\L,-2*\r) arc[start angle=270, end angle=90, radius=\r];
  \path (-0.65,0) pic[rotate=-90, scale=1.25] {aircraft};
  \draw (0,-\r) -- (\r,-\r);
  \node[font=\scriptsize, above] at (\r/2,-\r) {$r$};
  \node[fx] at (0,0) {};
  \node[above right, inner sep=2pt] at (0,0) {\textsc{fix}};
  \draw[dim] (-\L-\r-0.35,-\g) -- (-\L-\r-0.35,-2*\r+\g);
  \node[anchor=east] at (-\L-\r-0.4,-\r) {$2r$};
  \draw[dim] (\r+0.35,-\g) -- (\r+0.35,-2*\r+\g);
  \node[anchor=west] at (\r+0.4,-\r) {\texttt{maxXte}};
  \node[align=center] at (-\L/2,1.0)
    {\textbf{(a) the specification's speed}\\[-1pt]230 kt};
  \node[align=center] at (-\L/2,-2*\r-0.75)
    {$\texttt{maxXte}/2r = 1.00$\\[-1pt]inside the $\pm6\%$ band};
\end{scope}
\draw[dashed] (2.6,-2.9) -- (2.6,1.4);
\begin{scope}[shift={(7.9,0)}]
  \def\r{0.57}\def\L{2.44}
  \draw[crs] (-\L-\r-0.55,0) -- (\r+0.55,0);
  \draw[trk,-{Stealth[length=2mm]}] (-\L,0) -- (-\L/2,0);
  \draw[trk] (-\L/2,0) -- (0,0);
  \draw[trk] (0,0) arc[start angle=90, end angle=-90, radius=\r];
  \draw[trk,-{Stealth[length=2mm]}] (0,-2*\r) -- (-\L/2,-2*\r);
  \draw[trk] (-\L/2,-2*\r) -- (-\L,-2*\r);
  \draw[trk] (-\L,-2*\r) arc[start angle=270, end angle=90, radius=\r];
  \path (-0.65,0) pic[rotate=-90, scale=1.25] {aircraft};
  \node[fx] at (0,0) {};
  \node[above right, inner sep=2pt] at (0,0) {\textsc{fix}};
  \draw[dim] (-\L-\r-0.35,-\g) -- (-\L-\r-0.35,-2*\r+\g);
  \node[anchor=east] at (-\L-\r-0.4,-\r) {$2r'$};
  \draw[dim] (\r+0.35,-\g) -- (\r+0.35,-2*\r+\g);
  \node[anchor=west] at (\r+0.4,-\r) {\texttt{maxXte}$'$};
  \node[align=center] at (-\L/2,1.0)
    {\textbf{(b) the mutant's speed}\\[-1pt]200 kt, radius 24\% smaller};
  \node[align=center] at (-\L/2,-2*\r-0.95)
    {$\texttt{maxXte}'/2r' = 1.00$\\[-1pt]inside the $\pm6\%$ band\\[-1pt]
     \emph{both sides moved by} $0.76$};
\end{scope}
\end{tikzpicture}
\caption{The paper's central instance, drawn to scale. A holding pattern is a
racetrack an aircraft flies repeatedly over a fix; its across-course width is
exactly twice the turn radius $r$, so the maximum cross-track excursion
\texttt{maxXte} measures that width. A speed-table mutation taking 230 to 200
knots shrinks the bank-limited radius by $(200/230)^2 = 0.76$, and panel (b) is
drawn at that scale. The aircraft is the same size in both panels, because the
mutation shrinks the pattern and not the aeroplane. The result is visibly wrong,
and 24\% is far outside the $\pm 6\%$ band the property test allows.
The oracle passes anyway, having sized its band by calling the same turn-radius
function at the same wrong speed. Section~\ref{sec:measuring} measures what that
costs.}
\label{fig:hold-geometry}
\end{figure}
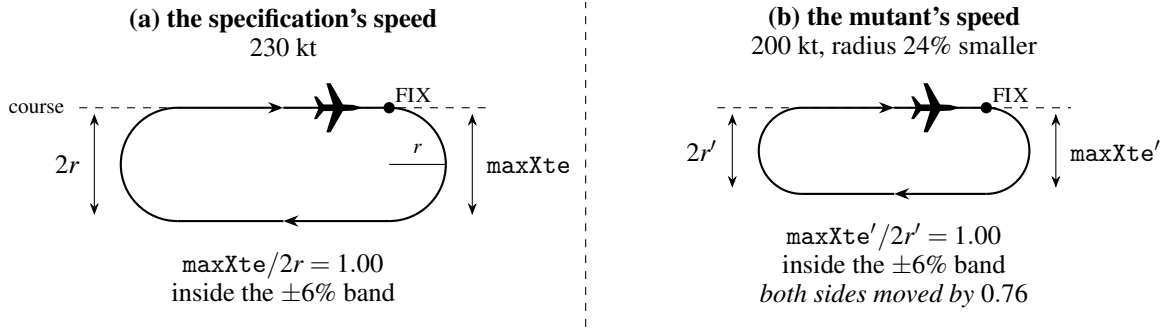

Two production defects in our simulator motivated the work reported here, one
week apart, both at the same waypoint on the same arrival. In the first, raw
approach eligibility flickered every frame while an aircraft flew a base leg. The
data tag never escalated and the aircraft rejected clearances a controller could
see it should accept. The remedy is a \emph{debounce}, named for the way a
mechanical switch contact bounces and reports several presses for one. It holds a
flickering value steady across brief dropouts and releases it only once the signal
has stayed away long enough to mean it. In the second, an aircraft assigned a hold flew a stable,
correctly shaped racetrack 8.5 nautical miles (NM) from the fix it was
supposed to orbit, and never crossed it. A \emph{holding pattern} is a racetrack an
aircraft flies repeatedly over a fix, so orbiting the right shape beside the
wrong point is a complete failure of the procedure, and
Figure~\ref{fig:hold-geometry} draws the geometry this paper needs from it. In
both cases the example tests covering the code passed. In the second case a
dedicated property suite for the holding state machine also passed throughout,
because every oracle in it concerned the sequence of legs and none concerned
where the aircraft was. We responded by adopting invariant-first, model-free
property testing for temporal behavior, and by writing 12 such suites.

This paper does not argue that this was the right response. It asks the question
the suites themselves cannot answer. \textbf{Do these oracles detect anything?}
The question is not rhetorical. A property test that passes tells you nothing
about whether it \emph{would} fail. Our discipline included a manual check in
which each oracle was run against a hand-picked one-line mutation of the code it
judges, but a hand-picked mutation is chosen by the same person who wrote the
oracle, and it therefore inherits that person's blind spots. Mutation testing
\cite{demillo1978,jia2011} replaces the hand-picked mutation with a mechanically
generated population, and mutation analysis of the resulting kill sets is how we
measure these oracles.

The answer we measured is unflattering and, we will argue, structural in origin. Across 4 modules and 366 distinct mutants, the invariant suites
kill 3 mutants that the pre-existing hand-written tests do not. The
explanation is not that the properties were written carelessly. It is that
several of them obtain the value they assert against from the very code they are
judging, so a fault moves both sides of the comparison at once and the assertion
remains true by construction. We call this property of an oracle its
\emph{anchoring}.

Anchoring is binary, decidable without running anything, and relative to a declared
mutate target. An oracle is \emph{specification-anchored} when its expected value is
composed from constants, published procedures or values fixed outside that target,
and \emph{state-anchored} when the expected value flows, directly or transitively,
from the code being mutated.

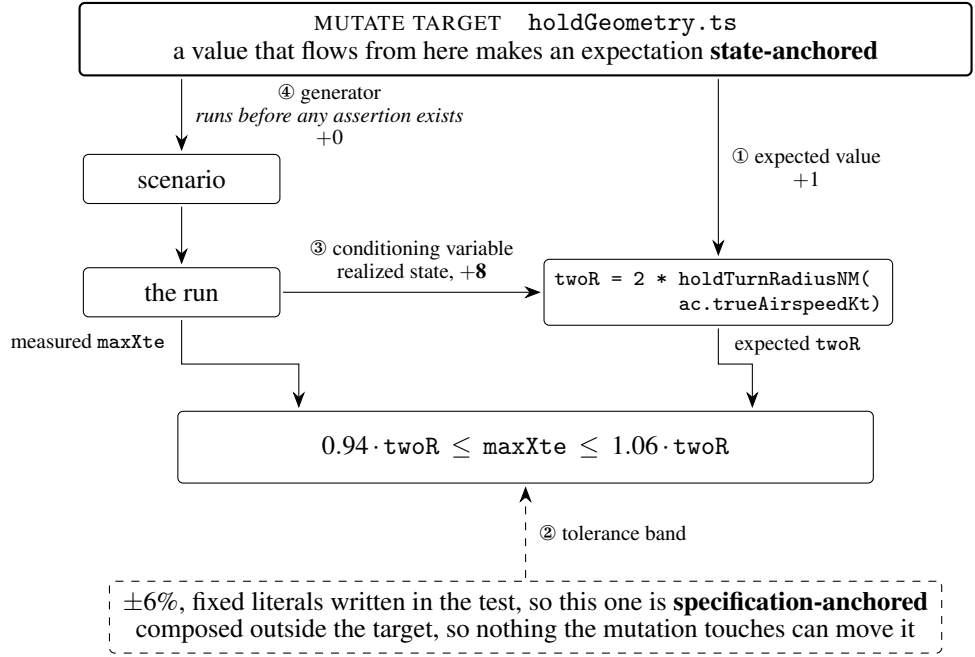
\begin{figure}[t]
\centering
\begin{tikzpicture}[
  font=\footnotesize,
  box/.style={draw, rounded corners=2pt, minimum height=6.5mm, inner xsep=5pt, align=center},
  mut/.style={box, thick},
  code/.style={box, font=\ttfamily\scriptsize, align=left, inner sep=4pt},
  off/.style={box, dashed, align=center},
  flow/.style={-{Stealth[length=2.2mm]}, shorten >=1pt, shorten <=1pt},
  dead/.style={-{Stealth[length=2.2mm]}, dashed, shorten >=1pt, shorten <=1pt},
  ch/.style={font=\scriptsize, inner sep=2pt, align=center},
]

\node[mut, minimum width=118mm] (mt) at (0,5.9)
  {\textsc{mutate target}\quad\texttt{holdGeometry.ts}\\[-1pt]%
   a value that flows from here makes an expectation \textbf{state-anchored}};

\node[box, minimum width=26mm] (scen) at (-4.55,4.05) {scenario};
\node[box, minimum width=26mm] (run)  at (-4.55,2.55) {the run};

\node[code] (exp) at (2.55,2.55)
  {twoR = 2 * holdTurnRadiusNM(\\\phantom{twoR = 2 * }ac.trueAirspeedKt)};

\node[box, minimum width=92mm, minimum height=9.5mm] (as) at (0,0.5)
  {$0.94\cdot\texttt{twoR} \;\le\; \texttt{maxXte} \;\le\; 1.06\cdot\texttt{twoR}$};

\node[off, minimum width=70mm] (band) at (0,-1.75)
  {$\pm 6\%$, fixed literals written in the test, so this one is
   \textbf{specification-anchored}\\[-2pt]
   composed outside the target, so nothing the mutation touches can move it};

\draw[flow] (mt.south -| scen.north) -- (scen.north)
  node[ch, midway, right=3pt] {\chnum{4}\;generator\\[-1pt]
    {\itshape runs before any assertion exists}\\[-1pt]$+0$};
\draw[flow] (mt.south -| exp.north) -- (exp.north)
  node[ch, midway, right=3pt] {\chnum{1}\;expected value\\[-1pt]$+1$};

\draw[flow] (scen) -- (run);
\draw[flow] (run.east) -- (exp.west)
  node[ch, midway, above=2pt] {\chnum{3}\;conditioning variable\\[-1pt]
    realized state,\ $\mathbf{+8}$};

\path (run.south) -- (-4.55,1.6) node[ch, midway, left=4pt] {measured \texttt{maxXte}};
\draw[flow] (run.south) -- (-4.55,1.6) -| ([xshift=-30mm]as.north);
\path (exp.south) -- (2.55,1.6) node[ch, midway, right=4pt] {expected \texttt{twoR}};
\draw[flow] (exp.south) -- (2.55,1.6) -| ([xshift=30mm]as.north);

\draw[dead] (band.north) -- (as.south)
  node[ch, midway, right=3pt] {\chnum{2}\;tolerance band};
\end{tikzpicture}
\caption{The four channels, located in the study's costliest oracle. Three lie in
the assertion and one upstream of it. Two share a single line. The \emph{yardstick} \texttt{twoR}, the value the
assertion compares the measurement against, is computed by the module under
mutation (\chnum{1}) and evaluated
at the aircraft's realized true airspeed (\chnum{3}). Closing the first recovers
1 mutant, the second 8, which is why naming the defect and knowing what it costs
are different things. The tolerance band (\chnum{2}) is composed from literals
fixed in the test, so nothing flows to it from the target. It is the one channel
of the four for which we found no anchored instance in any of the 12 suites, and
Table~\ref{tab:prevalence} carries the recovery figures.}
\label{fig:channels}
\end{figure}

A value reaches an oracle by one of four routes, which we call \emph{channels}.
Three lie in the assertion and one lies upstream of it: the \textbf{expected value}
the assertion compares against; the \textbf{width of a tolerance band}; a
\textbf{conditioning variable} the assertion is evaluated at; and a
\textbf{generator} that places the scenario, which runs before any assertion
exists. Only the first is named in the prior literature. The 4 are not conjuncts
of a compound assertion: an oracle makes one comparison, and the channels are four
ways the mutated code can reach into its terms. Figure~\ref{fig:channels} locates
all four in the holding suite, the study's costliest instance, where two of them
sit in a single line. Closing the expected-value channel recovers 1 mutant and closing the conditioning variable recovers 8 more, and the distance between naming this defect and knowing what it costs is what the rest of the paper is for. A mutant is \emph{recovered} when it survived the run before an intervention and is killed by the run after it, over the same population under the same pinned seed. Recovery is therefore a transition between two runs rather than a mutant status, and we report it by mutant identity rather than as a difference of two scores.

That measurement is where the study started and it is not what the paper is
about. A mutant is \emph{property-only} when the property suites kill it and the
hand-written tests do not. Counting those is how this paper measures what the
property suites \emph{add}, rather than what they detect on their own, and the
count runs one way: the reverse figure, mutants the hand-written tests kill
alone, is 64. A property-only count from one system, whose two suites were
written by the same person, is worth little on its own, and the closest published
comparison already reports the same direction over 40 projects
\cite{ravi2025}. What the rest of this paper does is intervene on anchoring instead of observing it: we remove it from one oracle and detection
rises, we insert it into a healthy one and detection falls, and we hold one oracle
fixed while moving only the declared mutate target and watch the same repair
become inert. Each effect was predicted before it was measured, and the prediction
was committed as its own revision beforehand, in order to maintain a strict
experimental methodology. It is the interventions, and not the property-only column,
that we ask a reader to weigh.

\paragraph{Research questions.} The measurement that started the study is RQ1,
and it is the least of the five. The paper is about RQ2 to RQ5.

\begin{description}\small
\setlength{\itemsep}{0pt}\setlength{\parsep}{0pt}\setlength{\topsep}{2pt}
\item[RQ1] Over identical mutant populations, what do the
      invariant property suites detect that the pre-existing hand-written example
      tests do not?
\item[RQ2] Where a property suite detects little, what is
      the cause, and is it a property of the oracle or of the code it judges?
\item[RQ3] Can that cause be intervened on, in both
      directions, with the effect predicted in advance?
\item[RQ4] How prevalent is it, and does its presence
      reliably cost detection?
\item[RQ5] Is anchoring a property of the oracle alone, or of
      the oracle together with the code under test?
\end{description}

RQ1 is answered in Section~\ref{sec:results}, RQ2 in Section~\ref{sec:anchoring},
RQ3 and RQ4 in Section~\ref{sec:measuring}, and RQ5 in Section~\ref{sec:scope}.

\paragraph{Contributions.} We claim the following, and we state at each point
what the prior literature already holds.

\begin{enumerate}
\item \textbf{A channel taxonomy and a provenance restriction.} The defect is
      already named at the expected-value channel, under at least 4 names
      \cite{garousi2018,soares2023catalog,rwemalika2021}, and separately as a
      soundness caveat in hardware assertion mining \cite{witharana2022}. We do
      not claim the name. We name the same defect at three further channels, a
      tolerance-band width, a conditioning variable read at evaluation time, and
      a scenario-placing generator, and we restrict the predicate to values that
      flow from the module under mutation, which is strictly narrower than
      "computed during the test" and is the version that determines killability.
\item \textbf{The first measurement of the cost.} Every prior source we
      retrieved motivates the smell by readability or maintainability. On one 46-mutant population,
      anchoring accounts for 9 mutants, 8 of them from a single instance.
      The channel the literature has named cost us nothing, because the
      hand-written tests happened to kill its one mutant anyway. The channel nobody
      has named is the only place on that module where the property suite catches
      something those tests miss.
\item \textbf{A bidirectional intervention.} Removing state anchoring recovers
      8 mutants on holding geometry; introducing it costs 4 on the
      debounce. On the same debounce population a reference-model suite kills
      exactly what the specification-anchored properties kill, which grounds the model-freedom argument in a measurement.
\item \textbf{Anchoring is scope-relative, and we measure it instead of asserting it.} One unchanged oracle, run against two declared mutate targets that
      differ only in whether they contain the code producing the value it reads,
      recovers nothing under one declared target and 3 mutants under the other. All 3 of
      the recovered mutants lie in that producer and none in the region the
      oracle exists to test. This also discriminates anchoring from the obvious
      rival explanation, propagation distance, which predicts no difference
      between the two cells because the oracle and the code producing its anchor
      are identically placed in both. Section~\ref{sec:distance} sets that
      comparison out.
\item \textbf{Anchoring is a provenance property that published oracle adequacy
      metrics cannot see.} Every metric surveyed by Hossain and Dwyer
      \cite{hossain2023}, the model-checking coverage taxonomy
      \cite{chockler2006fmsd}, and current hardware assertion-quality metrics
      \cite{witharana2022} measure forward influence. A self-referential oracle
      scores perfectly on all of them.
\item \textbf{A property-only result stated against a measured
      baseline.} 3 mutants of 366 as the suites were found, 4 once one of
      them is re-anchored. Ravi and Coblenz \cite{ravi2025} report
      the same direction over 40 Python projects, where unit tests outkill
      property tests roughly 3 to 1 in aggregate; where we differ is in
      \emph{overlap}, near-total here against 15\% of found mutations
      caught by both styles there. Per test our properties remain 6 to 33 times
      as efficient, so this is not a claim that property tests are weak, and any
      general subsumption claim is false in both datasets.
\item \textbf{A scope condition with an external consequence.} State anchoring is
      fatal only when the observed state and the mutated code are the same
      artifact. An oracle that reads realized positions from a \emph{separate}
      simulator is therefore unexposed to it \cite{li2024}, while ours is exposed,
      because our flight model and the code under test are one artifact.
\end{enumerate}

\section{Background and related work}

\paragraph{Oracle assessment by mutation testing.} That an oracle has a quality
independent of the code it exercises, and that mutation testing is the instrument for
measuring it, is established. Jahangirova et al.\ define oracle false
positives and false negatives and search for each
\cite{jahangirova2016,jahangirova2018tse}. Their false-negative criterion is our
diagnostic signature written out formally. The mutant is strongly killed, the
conditions in the assertions do not change value, and at least one variable
visible at the assertion point differs. OASIs packages the technique as a tool
\cite{jahangirova2018oasis}. Vera-P\'erez et al.\ partition undetected
mutants into no-infection, no-propagation and weak-oracle symptoms and measure
the split \cite{veraperez2019}. Terragni et al.\ automate the improvement
loop and impose a discipline we do not, using different mutation tools for
improvement and for validation to avoid circularity \cite{terragni2020}. Fraser
and Zeller \cite{fraser2012}, and Konstantinou et al.\ for large language
models \cite{konstantinou2024}, report the generated-oracle form of the same
failure, assertions that encode implemented rather than intended behavior. Huo
and Clause come closest to a provenance analysis, and they point the other way.
OraclePolish traces whether a value compared inside an assertion descends from an
input the test controls \cite{huo2014}. An oracle that is self-referential in our
sense is therefore well behaved in their terms, because its expected value
descends from a call the test itself makes. Barr et al.\ supply the taxonomy within
which the defect sits \cite{barr2015}. \emph{Mutation analysis as an
oracle-deficiency detector is not ours. What is ours is that the reaching half of
this defect class is decidable on the import-and-call graph, without executing
anything and without a mutant population, which is what every detector in this
cluster requires.} Only the flow half of the predicate is decidable; the sizing half, which
separates a costly instance from a free one in Section~\ref{sec:measuring},
required reading each oracle.

\paragraph{Test smells and tautological assertions.} This is the nearest prior
art and it owns the naming of our first channel. Garousi and K\"u\c{c}\"uk's
multivocal review records "the ugly mirror (also known as Tautological tests)"
\cite{garousi2018}. The Open Catalog of Test Smells carries \texttt{Self-Test},
\texttt{The Ugly Mirror} and \texttt{Test Tautology} as separate entries
\cite{soares2023catalog}. \texttt{Self-Test} is our expected-value channel
exactly, defined as tests that do not compare a result with an expected value but
with the result itself. Its worked example builds the expectation by calling a
different method of the same production class, which is the module-scoped form of
the predicate rather than a method-scoped one. Rwemalika et al.\ go furthest and give a computable
detector: the expected value should be a constant or a reference to a constant
and not computed during the test \cite{rwemalika2021}. None of these sources
connects the smell to fault detection, none runs mutation testing, and Rwemalika et al.\ state the impact of this smell as readability.

The shipping tools implement less than the prose. A rule-by-rule sweep of
\texttt{eslint-plugin-jest} \cite{eslintjest-artifact},
\texttt{eslint-plugin-vitest} \cite{eslintvitest-artifact}, tsDetect
\cite{tsdetect-artifact}, RTj \cite{rtj-artifact,martinez2020rtj} and SonarQube's
assertion rules \cite{sonarjs-artifact} found no implementation of the provenance
predicate; the Open Catalog's 518 defined entries ship documentation and no detector at
all. Every near miss decides inside the assertion expression, on either syntactic
identity of the operands or static decidability of the outcome. SonarJS rule
S5914 comes closest, and it comes at the question from the opposite side. It
reports a defect when ``the value is freshly created here, so the result is
independent of the code under test'', and it decides that by constant folding and
fresh-reference detection. Independence of the \emph{result} is its defect;
provenance of the \emph{expected value} is ours. The rule is therefore silent on
an expected value built by calling the module under mutation.

\paragraph{Tests that pass without discriminating.} The nearest published defect
class with a working detector is the complement of ours. A rotten green test
passes while containing an assertion that is never executed, detected by pairing
a static call-tree analysis with call-site-granular instrumentation
\cite{delplanque2019,aranega2021,googletest3811}. Vera-P\'erez et al.\
approach the same question from the production side with pseudo-tested methods
\cite{veraperez2018}, and Betka and Wagner give the closest thing to a taxonomy
of why a test failed to discriminate \cite{betka2021}. Zhang and Mesbah establish
that assertion quantity and assertion coverage correlate with mutation-measured
effectiveness \cite{zhang2015}. We find the converse. Our \emph{ablation} closes
one value flow at a time against a fixed mutant population, and it moves detection
from 26.09 to 45.65\% without adding an assertion. States A0 through A3a have
identical assertion coverage. \emph{Their defect
is an assertion that never runs; ours is an assertion that runs, reads the
infected state, and returns true anyway.}

\paragraph{Oracle adequacy metrics.} Hossain and Dwyer catalogue every
oracle-based adequacy metric published between 2007 and 2023, and find that all of
them share one direction \cite{hossain2023}. An element is covered when it is
executed and affects a value checked by an oracle through a dependency chain.
Schuler and Zeller's checked coverage is the sharpest of the family
\cite{schuler2011}. A self-referential oracle scores perfectly on all of them,
because the mutated function is in the dynamic slice of the assertion; it
influences both sides. The same direction was reached independently two decades
earlier in temporal-logic model checking \cite{chockler2006fmsd,hoskote1999} and
is current practice in hardware \cite{witharana2022,goldmine-artifact}. The one
prior provenance restriction we found, interface-aware signal temporal logic,
partitions a specification's signals by interface role, input against output
\cite{ferrere2019}, which is a different partition from ours and does not decide
killability.

\paragraph{Vacuity.} The closest mechanism-plus-detector analogue outside
software testing does not reach our defect. Vacuity detection quantifies over
subformulas of the specification and perturbs the specification
\cite{beer2001,armoni2003,kupferman2006,gheorghiu2006}; coverage in model
checking perturbs the system \cite{chockler2006fmsd,katz1999}. Our defect, a
relation between two quantities both produced by the mutated component, is
formally non-vacuous under every definition we retrieved, since replacing either
side by false falsifies the formula. The check does not merely miss it; it
certifies it healthy. Ball and Kupferman are the nearest miss, and they strengthen the
point \cite{ball2008}. They place a real oracle inside the vacuity framework and
define vacuity as branch coverage of it. Nothing there forbids that oracle from
calling the program to build its expectation, and in that case every branch is
still taken and the algorithm reports a clean pass. A paper that put an oracle inside the
vacuity framework and did not ask the provenance question is the best evidence
available that the question was not being asked.

\paragraph{Property-based testing, evaluated.} Ravi and Coblenz supply the only measured
baseline against which our property-only result can be read
\cite{ravi2025}. Over 40 Python projects, unit tests found 1{,}365 of 1{,}610
detected mutations against property tests' 467. They outkill roughly 3 to 1, and
do so in 32 of the 40 projects with 2 ties. The direction is ours. What differs is
overlap and efficiency. About 15\% of mutations were caught by both styles,
where ours nearly coincide. Property tests won outright in 6 projects, where ours
never do. Per test they are roughly 50 times as effective, the one axis on which
our suites behave like theirs. Their input sweep is the strongest
evidence against a "just run it more" remedy: 96\% of the mutations they
found came within 350 inputs. Goldstein et al.\
interview practitioners, 7 of 30 of whom name mutation testing as their check on
whether a property is real, just below example inspection at 8
\cite{goldstein2024}. Hughes's industrial
QuickCheck experience is the best-known report that property testing finds what
conventional testing misses, though it is an experience report rather than a
controlled evaluation \cite{hughes}. It also comes from a style we do not use. It
maintains an abstract model with state transitions per API operation, which is an
independent source of expected values and is therefore structurally incapable of
self-reference. Coverage-guided property testing
addresses the complementary problem of generator reach \cite{lampropoulos2019}.
Property-based testing descends from QuickCheck \cite{claessen2000} and
metamorphic testing from Chen et al.\ \cite{chen1998,segura2016}; both
appear in our suites.

\paragraph{Falsification oracles for cyber-physical systems.} Signal temporal
logic is specification-anchored in its semantics and state-anchored in its
practice. The robustness of an atomic proposition is the distance from the signal
to a set fixed by the formula \cite{abbas2014}, so anything state-anchored enters
through what people put inside that set. ST-Lib is the field's catalogue of what
a requirement looks like, and its named templates compare a signal against a
second signal or against the signal's own earlier value. Its authors state our
risk condition in their own words, that the expected steady-state value of the
signal may be unknown when the specification is created \cite{kapinski2016}. Requirement mining is the canonical case of a threshold
obtained from the system's own realized behavior \cite{jin2015}. The predicate
fires, unremarked, in the community's canonical benchmark. The ARCH-COMP air-fuel
control requirements bound a normalized quantity computed inside the model under
test \cite{archcomp-artifact,menghi2023}. Breach ships the same model, with its
requirement written against a reference signal rather than against that
normalized quantity \cite{breach-artifact}. We state that as a structural finding and a
hypothesis, not a measured detection loss, since we ran no mutation testing
there. Two sources bound it: Li et al.\ read realized positions safely
because plant and system under test are separate artifacts \cite{li2024}, which
in our simulator they are not; and Shin records that mutation-based adequacy
assessment for a whole simulated system does not yet exist \cite{shin2026}.
HECATE is the closest analogue to our test files \cite{formica2022}.

\paragraph{Mutation analysis of simulated and safety-relevant systems.} Mutation
analysis with simulation subjects exists: Bartocci et al.\ run it on an
aircraft elevator control system \cite{bartocci2023}, and Cornejo et al.\ at
212{,}502-mutant scale on satellite software tested through emulators
\cite{cornejo2022,vigano2022}. The air-traffic gap is real but thin, with one near
neighbor: Boukelloul et al.\ mutate agent coordination in an en-route air
traffic control model, with oracles anchored on published separation minima, and perform
no oracle analysis \cite{boukelloul2025}. Because the domain is thin, we rest no
novelty claim on the subject.

\paragraph{Fault propagation, and the explanation we rejected.} Voas's model
states that a fault is revealed only if it is executed, if it infects program
state, and if that infection propagates to an observable output \cite{voas1992}.
The distance an infection must travel to reach an oracle is the natural
explanatory variable to draw from that model, and it was ours until the scope
experiment separated it from anchoring. Section~\ref{sec:distance} sets out why we
rejected it, and records that the quantity is in any case already operationalized
by others \cite{niedermayr2019,veraperez2019,jahangirova2016}. Mutation analysis itself
originates with DeMillo et al.\ \cite{demillo1978} and is surveyed by Jia
and Harman \cite{jia2011}. Sargent's framework places this work as computerized
model verification \cite{sargent2013}.

\section{The system and its invariant suites}
\label{sec:system}

SECTOR is a deployed air traffic control simulator written in TypeScript. Its
users include a retired area controller who serves as the project's domain
reviewer, and the two defects described in Section~1 were both observed in live sessions; neither surfaced in testing. The subsystem relevant here is a deterministic
core containing the flight model, the command queue, the route and approach
controllers, conflict detection, and collision-avoidance logic. The core is free
of browser dependencies and uses a seeded pseudorandom generator, so a scenario
is reproducible from a seed. It is 17{,}688 lines across 60 files, within a
52{,}000-line application, and it is under continuous change: 29 commits landed on
it in the month during which the holding suite of Section~\ref{sec:results} ran
green on every one of them. The 4 modules we mutate are a small slice of it,
chosen for their structure; size was not a criterion. Every count in this
section is as of the commit the study ran against; the codebase has moved since.

The pre-existing hand-written tests are not a token baseline. The unit suite is
248 files and 3{,}819 tests, and it runs on every commit alongside a typechecking
gate. Every property-only count in this paper is measured against that
suite. Section~\ref{sec:results} reports that the property suites add little, and
this is what they add little to.

Twelve model-free invariant suites cover temporal behaviors in this core. They
contain 47 tests, of which 44 are generated properties written with
\texttt{fast-check}, and a full run executes 9{,}870 traces in about 1.4 seconds.
Two rules define their construction.

First, every oracle is either an invariant computed from the input trace together
with a named specification constant imported from the source, or a metamorphic
relation between two executions of the real code. None of the 12 maintains a
reference model. One file elsewhere in the tree does, a model-conformance
proof-of-concept for the debounce that predates the methodology, and
Section~\ref{sec:measuring} runs it as a control. The reasoning was that a reference model is a second implementation
that drifts as the real code changes and can carry the same fault as the code it
judges. That reasoning is sound, and Section~\ref{sec:anchoring} argues that it also
removes the only independent source of expected values, which is the condition under
which the defect we report becomes natural to write.

Second, because safety invariants are one-directional, and are satisfied by code
that does too little as readily as by code that is correct, every suite is
required to carry at least one oracle of the liveness, exactly-once, additivity
or determinism kind. In the motivating debounce defect the natural safety bound,
that eligibility is never held longer than the specified window, did not
discriminate, because the aircraft released eligibility too early.

The suites are not a continuous integration gate. They run as part of the
ordinary pre-commit test invocation and are treated as reference artifacts to be
re-derived when the behavior they describe is touched.

\paragraph{Where the constants come from.} The argument in
Section~\ref{sec:anchoring} turns on specification anchoring having been
available and free, so it matters which of the quantities this paper prints are
published and which are ours. Three are published. The holding turn is
constructed at the lesser of Rate One (3 deg/s) or a fixed bank angle, which is
the construction given by the International Civil Aviation Organization in its
Doc 8168 (PANS-OPS) \cite{skybraryhold} and by the US Federal Aviation
Administration in its Aeronautical Information Manual (AIM) 5-3-8j, the latter directing that all turns be made at ``3 degrees per
second, or 30 degree bank angle, or 25 degree bank angle, provided a flight
director system is used'', whichever requires the least bank \cite{faaaim2026}.
The maximum holding indicated airspeeds by altitude band (200, 230 and 265 kt),
and the 6{,}000 and 14{,}000 foot breaks between those bands, are AIM 5-3-8j as
well \cite{faaaim2026}. Those bands are the FAA's: ICAO publishes different ones,
so this is a US convention applied uniformly across our airports. The vertical
separation minima are ICAO Doc 4444 (PANS-ATM): 1{,}000 ft below flight level
(FL) 290,
1{,}000 ft between FL 290 and FL 410 inclusive where the reduced vertical
separation minimum (RVSM) is in force, and
2{,}000 ft above FL 410 \cite{icaorvsm2002}. Our implementation is that rule,
which is why its constant is named for the RVSM ceiling rather than for FL 290.

Three others that this paper prints are modeling choices, and we mark them as such instead of letting them sit undifferentiated beside the
published values. The density model \texttt{estimateTAS} is ours. So is the
four-second eligibility hold the debounce implements, which was tuned to bridge
geometric seams and has no external referent. And so, importantly, is the
two-nautical-mile horizontal trigger for a resolution advisory: a real traffic
alert and collision avoidance system (TCAS) II
triggers on projected time to closest approach with a horizontal threshold that
varies by sensitivity level, not on a fixed range test, so the flat 2 NM used here
is a simplification. It is adequate for the behavior under study, which concerns what an advisory does
once triggered; the fidelity of the trigger is out of scope. It is not a standard,
and nothing in this paper should be read as claiming it is.

\section{Study design}
\label{sec:design}

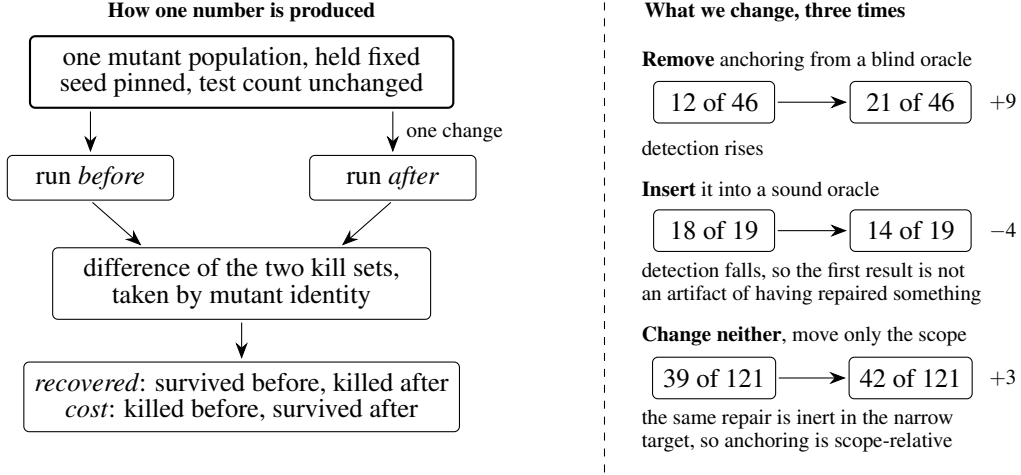
\begin{figure}[t]
\centering
\begin{tikzpicture}[
  font=\footnotesize,
  box/.style={draw, rounded corners=2pt, minimum height=5.5mm, inner xsep=4pt, align=center},
  pop/.style={box, thick},
  flow/.style={-{Stealth[length=2.2mm]}, shorten >=1pt, shorten <=1pt},
  lbl/.style={font=\scriptsize, inner sep=2pt, align=center},
  txt/.style={font=\scriptsize, inner sep=1pt, anchor=west, align=left},
]
\node[lbl, font=\scriptsize\bfseries] at (-4.3,3.95) {How one number is produced};
\node[pop, minimum width=56mm] (pop) at (-4.3,3.15)
  {one mutant population, held fixed\\[-2pt]seed pinned, test count unchanged};
\node[box, minimum width=22mm] (bef) at (-6.3,1.75) {run \emph{before}};
\node[box, minimum width=22mm] (aft) at (-2.3,1.75) {run \emph{after}};
\draw[flow] (pop.south -| bef.north) -- (bef.north);
\draw[flow] (pop.south -| aft.north) -- (aft.north)
  node[lbl, midway, right=3pt] {one change};
\node[box, minimum width=50mm] (dif) at (-4.3,0.35)
  {difference of the two kill sets,\\[-2pt]taken by mutant identity};
\draw[flow] (bef.south) -- ([xshift=-13mm]dif.north);
\draw[flow] (aft.south) -- ([xshift=13mm]dif.north);
\node[box, minimum width=56mm] (out) at (-4.3,-1.15)
  {\emph{recovered}: survived before, killed after\\[-2pt]%
   \emph{cost}: killed before, survived after};
\draw[flow] (dif.south) -- (out.north);

\draw[dashed] (0.5,-2.15) -- (0.5,4.15);

\node[lbl, font=\scriptsize\bfseries, anchor=west] at (0.95,3.95) {What we change, three times};
\def\row#1#2#3#4{%
  \node[box, minimum width=16mm] (r#1a) at (1.95,#2) {#3};
  \node[box, minimum width=16mm] (r#1b) at (4.55,#2) {#4};
  \draw[flow] (r#1a) -- (r#1b);
}
\node[txt] at (0.95,3.30) {\textbf{Remove} anchoring from a blind oracle};
\row{1}{2.75}{12 of 46}{21 of 46}
\node[txt] at (5.55,2.75) {$+9$};
\node[txt] at (0.95,2.15) {detection rises};

\node[txt] at (0.95,1.60) {\textbf{Insert} it into a sound oracle};
\row{2}{1.05}{18 of 19}{14 of 19}
\node[txt] at (5.55,1.05) {$-4$};
\node[txt] at (0.95,0.35) {detection falls, so the first result is not\\[-1pt]an artifact of having repaired something};

\node[txt] at (0.95,-0.35) {\textbf{Change neither}, move only the scope};
\row{3}{-0.90}{39 of 121}{42 of 121}
\node[txt] at (5.55,-0.90) {$+3$};
\node[txt] at (0.95,-1.60) {the same repair is inert in the narrow\\[-1pt]target, so anchoring is scope-relative};
\end{tikzpicture}
\caption{The measurement, and the three interventions built on it. No number here
is a score on its own. Each is the difference between two runs over one fixed
population with the seed pinned and the test count unchanged, taken by mutant
identity rather than by subtracting scores, because two runs can kill the same
number of mutants and kill different ones. Each intervention changes exactly one
thing and reports that difference. The holding figures are A0 and A3a of
Table~\ref{tab:ablation}, so the 9 is what closing every anchoring channel on
that module recovers; 8 of the 9 come from the single conditioning instance, and
Section~\ref{sec:measuring} draws that distinction once. The scope
intervention is the S2 pair of Table~\ref{tab:scope}.}
\label{fig:method}
\end{figure}

We selected four targets spanning the range of structural relationships between
an oracle and the code it judges.

\begin{itemize}
\item \textbf{Holding geometry.} A 90-line module computing holding turn rate,
      turn radius, target speed and outbound leg time, drawn to scale in
      Figure~\ref{fig:hold-geometry}. Covered indirectly by two suites, one over
      the holding state machine and one over the flown trajectory. This is the
      paper's central instance and the one Figure~\ref{fig:channels} anatomizes.
\item \textbf{Debounce.} Lines 97 to 118 of the route-following controller, the
      eligibility hold that motivated the methodology. The signal it steadies is
      approach eligibility, a geometric test that can go false for an instant
      without anything having changed, and the hold releases after four seconds. The covering property
      suite asserts a bound on the hold duration, a no-resurrection property, a
      time-step additivity relation, and determinism.
\item \textbf{Angle and frame arithmetic.} A 61-line module of pure functions for
      angle normalization, shortest angular delta, bearing, magnetic and true
      frame conversion, and planar distance. Twelve metamorphic properties cover
      it directly.
\item \textbf{Collision-avoidance controller.} A 309-line module for resolution
      advisory triggering, sense resolution and phase management. Covered by a
      suite asserting that every controller command is rejected while an advisory
      is active and that the active phase persists for a minimum duration.
\end{itemize}

A fifth target enters later and only for the scope experiment of
Section~\ref{sec:scope}: the minimum safe altitude warning (MSAW) block inside the physics loop,
together with the vertical-speed computation in the same file. It takes no part in
the comparison of Section~\ref{sec:results}, because there is no hand-written
example suite over that block to compare against, and reporting it in
Table~\ref{tab:main} would mean inventing a baseline. It is used for what it is
uniquely suited to, which is varying the mutate scope under a fixed oracle.

For each of the four comparison targets we performed two mutation testing runs with Stryker, using the
\texttt{vitest} runner, per-test coverage analysis, a concurrency of 6, and a
30-second timeout. The first run executed only the invariant properties covering
the target. The second executed only the hand-written example tests covering it,
excluding property tests. Restricting the test set per run is what makes the
comparison possible: a run over the whole suite measures the project, not the
style.

\paragraph{The partition is derived from coverage, not from imports.} Our first
attempt assigned test files to the two categories by reading their imports. That
specification is wrong, and it is the single largest error we made in this study.
The hand-written example suite for the debounce never
imports the route-following controller; it drives the debounce through the
physics update. Imports are syntactic; coverage is empirical. The partition
reported here is derived from Stryker's own \texttt{coveredBy}: each target is
run once against the whole unit suite, each mutant's covering test identifiers
are mapped back to files, and every file covering at least one mutant enters the
partition. Under the import rule the debounce example suite appeared to score
21.05\% with no coverage at all on two branches; under the coverage rule it
scores 89.47\%, and the "no coverage" finding was an artifact of the partition. We report the corrected figures
throughout. It is also the reason this paper makes no complementarity claim: under
the import rule the two styles looked complementary, and under the coverage rule
they very nearly coincide.

One hand rule survives on top of the derived partition and we state it here. Files matching \texttt{*.property.test.ts} are
excluded from both buckets, because counting a generator-driven model-conformance
suite as an example test would credit the hand-written style with property-test
detection, and counting it as one of our invariant suites would credit the
model-free discipline with a model-based result. Exactly one file falls in that
bucket, and instead of letting its exclusion stand as an undisclosed convenience we
run it separately in Section~\ref{sec:measuring} against the same mutant
population.

\paragraph{The mutate scope is a parameter, and the debounce differs from the
others.} Three of the four targets mutate a whole file. The debounce mutates
lines 97 to 118 of a 604-line file, which matters because the constant its oracles
assert against sits at line 37 of that same file: widen the scope to the whole
file and those oracles would become state-anchored by our own definition. Two
things bound this. The predicate is decidable relative to a \emph{declared} mutate
scope, not absolutely, and we say so wherever it is stated. And the range was not
drawn to produce the result: the debounce runs are timestamped more than 8 hours before the first ablation run and nearly 10 before the negative control, so
the scope predates the anchoring thesis rather than following from it. A reader
with no mutant population has no mutate target either, and
Section~\ref{sec:implications} says what the predicate reduces to in that case.

\paragraph{Mutant identity and denominators.} Every number this paper reports is
a difference between two runs taken on this basis, which
Figure~\ref{fig:method} draws. Mutants are matched across runs on
the tuple (start line, start column, mutator name, replacement text). That tuple
is not unique: a few mutants share it, and where they do their statuses can
disagree, so collapsing them with a last-write-wins map yields an order-dependent
answer. We report raw and distinct counts separately and resolve conflicts under
an explicit rule, \emph{any-killed}, meaning an identity counts as detected if
any raw mutant carrying that key was killed, which is the semantically correct
rule for the question "does this suite detect this change." The alternative rule
is computed alongside it, and the difference is confined to how many mutants are
credited to both suites at once: 3 on the collision module and, in the scope
experiment of Section~\ref{sec:scope}, one and two. \emph{No property-only column
in this paper moves under either rule}, which matters because those columns are
what every conclusion here is drawn from. We assert both rules in the validation program instead of reporting the one that reads better. Scores use Stryker's documented default denominator,
$(\mathrm{Killed}+\mathrm{Timeout})/(\mathrm{Killed}+\mathrm{Timeout}+\mathrm{Survived}+\mathrm{NoCoverage})$,
so uncovered mutants are inside the denominator and timeouts count as killed
\cite{strykerdocs}. Scores are over raw mutants; the survivor-diff columns are
over distinct identities, which is why the two sets of totals differ slightly on
two modules. No retrieved study shares this identity key, and few share the
denominator, so Section~\ref{sec:threats} states both conventions.

\paragraph{Equivalent mutants.} Equivalent mutants, which change no behavior and
cannot be killed, inflate the apparent weakness of any suite. We classify by hand and report explicitly instead of adjusting scores. The angle module and the
debounce are classified exhaustively, at two and one respectively. The 165
collision mutants surviving both suites are not: they are bounded instead by a
30-mutant sample drawn from a recorded seed, reported in
Section~\ref{sec:results}. The comparable cyber-physical literature strips 17 to
22\% of all mutants as equivalent \cite{bartocci2023}, and one such study strips
42\% of the \emph{live} mutants \cite{cornejo2022}, so our absolute percentages read low against
that convention for two independent reasons, an inflated denominator and an
unclassified equivalent fraction.

\paragraph{Reproducibility.} Every quantity in this paper is emitted by one
program, which reads the raw Stryker reports and computes scores, diffs,
per-function breakdowns and the published dataset. Nine figures we had computed by
hand off a terminal were recomputed against it, and the tool was right in every case
of disagreement. It also carries a gate asserting every mutation figure printed here against
the report it came from, under both conflict rules, and exiting non-zero on any
mismatch. We added the gate late: the version it replaced pointed at superseded
import-derived reports and at the wrong ablation state, so it agreed with itself
while disagreeing with the draft on 5 figures, and it printed a table rather than
failing. A reproducibility artifact that cannot fail is the same defect this paper is
about, one level up.

\paragraph{Use of AI tools.} Claude Opus 5, accessed through the Claude Code
command-line tool in August 2026, wrote the analysis program that emits every
figure reported here, its validation gate and the reproduction script published
with the dataset, and drafted and revised the text of this manuscript under the
author's direction, including two restructurings of its argument. A coding agent
wrote the ablation repairs and the state-anchored negative control to
specifications the author fixed in advance, which is part of the study design. Claude Opus 4.8 and Claude Opus 5 performed
the prior-art retrieval and the adversarial review whose challenges are
dispositioned throughout. The mutated modules and the invariant suites under
evaluation were authored the same way, which Section~\ref{sec:threats} treats as a
threat rather than a disclosure. The author directed and verified all of it and
takes full responsibility for the content; no AI technology is an author. No
quantity reported here was produced by a language model: every figure is computed
by \texttt{analyze.ts} from the raw reports, and every mutation figure is asserted
against them by a gate that exits non-zero on any mismatch.

\section{Results}
\label{sec:results}

Table~\ref{tab:main} reports the mutation score for each suite run in isolation,
for the two run together, and the survivor diff between them.

\begin{table}[t]
\centering\footnotesize
\begin{tblr}{colspec={lrrrrrrrr}, row{1}={font=\bfseries}}
\toprule
Module & Raw & Dist. & Prop. & Ex. & Stacked & P-only & E-only & Neither \\
\midrule
Holding geometry    & 46  & 44  & 12 (26.09\%)            & 35 (76.09\%) & 35 (76.09\%)$^{\ast}$ & 0 & 21 & 11  \\
Debounce            & 19  & 19  & 18 (94.74\%)            & 17 (89.47\%) & 18 (94.74\%)          & 1 & 0  & 1   \\
Angle arithmetic    & 50  & 50  & 40 (80.00\%)            & 48 (96.00\%) & 48 (96.00\%)          & 0 & 8  & 2   \\
Collision avoidance & 264 & 253 & 54 (20.45\%)$^{\dagger}$ & 90 (34.09\%) & 92 (34.85\%)          & 2 & 35 & 165 \\
\midrule
\SetCell[c=9]{l}\emph{after the oracle repair of Section~\ref{sec:measuring}, reported here so the
headline is not computed from it:} & & & & & & & & \\
Holding geometry    & 46  & 44  & 21 (45.65\%)            & 35 (76.09\%) & 36 (78.26\%)          & 1 & 13 & 10  \\
\bottomrule
\end{tblr}
\caption{Mutants killed, with the score in brackets, for each suite run in isolation and for the two run together (``Stacked''), then mutants killed by exactly one of the two, under a coverage-derived partition. Counts are the numbers the prose quotes. Rows above the rule are the suites \emph{as found};
the row below is the same population after Section~\ref{sec:measuring}'s
re-anchoring, an intervention rather than an observed state. Scores are over raw
mutants, diff columns over distinct identities, which is why the totals differ
where they differ. $^{\ast}$The one stacked figure not separately executed: the
as-found holding property suite kills a strict subset of the example suite's
kills, so the stack is the example run. $^{\dagger}$Collision figures are lower
bounds: a 30-mutant sample of the 165 both-survivors classifies 6.7\%
unkillable, Wilson 95\% [1.8, 21.3], placing the true property score between
roughly 20.5 and 24\%.}
\label{tab:main}
\end{table}

\paragraph{The headline is the property-only column, and it is small.} Summed across
the 4 modules, the property tests covering them kill 3 mutants out of 366
that the pre-existing hand-written tests do not: one, zero, zero and two. We state
that number as found rather than as repaired, because the fourth property-only mutant
exists only after an intervention this study performed. Re-anchoring the holding
oracle in Section~\ref{sec:measuring} raises that module's property-only count
from zero to one and the total from 3 to 4, and every later reference to
``4'' in this paper carries that condition. The unit measured is the property
tests covering these 4 modules, which is not a clean subset of the 12 suites:
\texttt{math.ts} is shared, so its coverage-derived partition draws tests from 11
of them. No claim here extends to what those suites cover on modules we did not
mutate. Two of the 3 non-zero
contributions in the repaired total are exact boundary values that a hand-written
case did not happen to pin. On the debounce it is the release comparison mutated
from \texttt{>=} to \texttt{>}. On holding geometry it is the altitude band
boundary, \texttt{altitudeFt <= 14000} mutated to \texttt{<}, which the property
generator reaches because it draws altitudes as thousands and therefore lands on
14{,}000 exactly. This is the classic argument for generated tests, and it is
worth exactly two mutants here. The third non-zero contribution is the pair on
collision avoidance, which are not boundary values.

\paragraph{Stacking the two styles is what a project actually does, and it buys
almost nothing.} Few projects ship one style, so we ran the union as well, on every
module. Against the better of the two halves it recovers zero, zero, zero and two
mutants, and one more on holding geometry once that suite is re-anchored. On the
angle module the stack is indistinguishable from the example suite alone and on the
debounce from the property suite alone. Ravi and Coblenz report the same aggregate
direction over 40 Python projects, though with far less overlap than we find
\cite{ravi2025}.

That is worth one methodological remark. \emph{Composition is exact.} On all 4
modules the union kills precisely the union of what the two suites kill in
isolation, to the individual mutant, and the both-survivor counts are identical in
every view. Tests share process state and run in an order, so two suites measured
apart need not compose when run together; here they do, so every stacked figure
above is recoverable from the published dataset without the four extra runs. We executed them instead of deriving them because deriving across the raw and distinct
denominators is the error that damaged an earlier pass of this study.

\paragraph{Per test, the properties are far more efficient, and that matters for
how this result should be read.} Dividing kills by the number of tests in each
partition gives 4.50 against 0.21 on the debounce, 0.93 against 0.03 on the angle
module, 1.00 against 0.18 on holding geometry, and 27.00 against 2.57 on collision
avoidance, so the property side is between 6 and 33 times as efficient per test.
The holding figure is the suite as found; re-anchored it kills 21 rather than 12
over the same 13 tests, which would read 1.62 and 9-fold.
The denominator is unflattering to the example side, because a coverage-derived
partition sweeps in files that touch the module incidentally, 84 files and 1{,}708
tests in the angle case, so these ratios are an order of magnitude rather than a
measurement. Read with that caveat they still matter: our result is that two
styles nearly coincide in what they detect, not that generated properties are a
poor use of effort. Ravi and Coblenz report roughly fiftyfold on the same axis
\cite{ravi2025}, so this is the one dimension on which our suites behave like
theirs.

\paragraph{Holding geometry.} As found, the property suite killed nothing the
example tests did not already kill, at 26.09\% against 76.09. That is the
result which prompted the rest of this paper. Its 32 distinct survivors are the
21 the example suite kills alone plus the 11 both suites miss, and they divide
cleanly: 10 in speed selection, 15 in two functions the harness never calls, 4 in
leg timing behind a lap-duration bound carrying several minutes of deliberate
slack, 2 in turn radius and 1 in turn rate. The 10 in speed selection are the
important group, and the reason they survive is not that no oracle observes their
effect. It is that the geometric oracles are conditioned on the aircraft's
realized true airspeed, so a wrong speed produces a correctly shaped racetrack
\emph{for that wrong speed}. Section~\ref{sec:anchoring} develops that sentence
into the paper's thesis, and Section~\ref{sec:measuring} shows that 8 of the
10 die once the yardstick is re-anchored.

It is worth saying explicitly what anchoring does not explain here, because the
repair nearly doubles the score and that invites over-attribution. Anchoring owns
9 of the 32 survivors. The largest single group, the 15 in two functions the
harness never calls, is a generator and harness limitation with no oracle content
whatever, and is the production-side blind spot Vera-P\'erez et al.\ measure as
pseudo-tested methods \cite{veraperez2018}, and it is also why this module is the one most sensitive to the
denominator convention: excluding uncovered mutants moves the as-found figure from
26.09 to 41.38\%. The repair nearly doubles the score under either
convention, 26.09 to 45.65 or 41.38 to 72.41, but the mechanism this paper is about accounts for 9 mutants; it does not account for the whole distance between the two suites.

\paragraph{Debounce.} The property suite kills 18 of 19 and the example suite 17
of 19, and the properties are very nearly a subset rather than a superset. The
single property-only kill is the release boundary described above. The single
mutant surviving both negates a guard whose false branch is unreachable given a
call-site invariant documented on the method, and is equivalent. This module is
the reference case for a healthy oracle, and Section~\ref{sec:measuring} uses it
as such: its assertions compare against a constant imported from outside the
mutated line range, which no mutation in the population can perturb.

\paragraph{Angle arithmetic.} The example suite kills 48 of 50, and the two
survivors are the two we argue are equivalent: a comparison whose alternative
branch is absorbed by a downstream guard, and an addition of 180 degrees mutated
to a subtraction, which normalizes identically. The example suite therefore kills
every killable mutant in the module, and there was no headroom for the property
suite to contribute. That zero is a ceiling effect rather than a weakness in the
properties, and it is the clearest illustration of why a property-only
count must be read against the population it is computed over.

\paragraph{Collision avoidance.} Both suites are weak and 165 of 253 distinct
mutants survive both, so every collision figure in this paper is a bound: 30 of
those 165 were sampled for equivalence and 135 remain individually unclassified. The surviving population is concentrated where the
advisory's content is decided rather than where its contract is: 56 in sense
resolution and direction scoring, 37 in the update and trigger path, and 21 in
parallel-approach detection. The property suite is sound for what it claims,
which is the advisory's contract with the controller while active, namely that
every controller command is rejected for as long as an advisory runs and that the
phase persists for its minimum duration. It claims nothing about whether an
advisory should have fired at all, or which aircraft is sent which way.
A \emph{sense} is the direction of a resolution advisory, climb or descend.
Coordinated opposite senses is the defining property of the collision-avoidance
concept, and no oracle in either suite constrains it. That is a gap in the suite and not a fact about property testing, and it is the one module where we
would expect a competent second author to do materially better.

Because 165 unclassified survivors invite the worst available reading, we sampled
30 of them from a recorded seed and argued each individually: 2 are behaviorally
inert and 28 are genuine undetected behavior changes, an unkillable proportion of
6.7\% with a Wilson 95\% interval of [1.8, 21.3] that puts the true
property score between roughly 20.5 and 24\%. The bound supports the
conclusion the raw number suggests: this module genuinely is close to untested by
both suites, and saying so with an interval is both more accurate and more
defensible than a point estimate.

One reading it does not support is that the module is close to broken. A surviving
mutant establishes only that no oracle in either suite pins the behavior that
mutant changes, and every figure in this paper measures oracles and not the correctness of the code they run against.

The classification pass is worth reporting in one further respect, because it
repeated an error the rest of this paper is about. A first pass put 4 of the 30
in a separate class, differing from the original only at an exact value that a
continuous scenario space would not produce, and counted them as unkillable at
20.0\%. That premise was wrong for this harness. Our scenarios do not
integrate positions from continuous inputs; the covering tests assign them directly
at round numbers, so an above-ground-level test is an integer difference and a
two-nautical-mile range is \texttt{hypot(2.0, 0)} exactly. The fourth was settled by
reading the closest-approach routine, where the time to closest approach is clamped
by a \texttt{max(0, \dots)} and is therefore exactly zero for every already-diverging pair, so the value never sits at a knife edge. The tell was available in our
own results, since the single property-only mutant this study reports on holding geometry
is itself an exact boundary hit.

Rather than reclassify on that reasoning alone we wrote the 4 killing tests, each
constructing the exact value and carrying an intermediate assertion showing it is
genuinely produced. All 4 pass against clean source, and run as the sole test set
all 4 mutants die, so these are gaps established by execution. The
tests were then discarded rather than adopted, because keeping them would close 4 real gaps at the cost of changing every collision figure reported here; that is a
decision for after the measurement rather than during it. Note the direction of the
whole correction: it tightened the interval and made this module look worse.

\section{Oracle anchoring}
\label{sec:anchoring}

The four measurements are not contradictory once one asks where each oracle
obtains the value it compares against, which is the anchoring defined in
Section~\ref{sec:intro}. We restate the distinction here because the rest of this
section is the argument that it is the operative one.

\begin{quote}
\textbf{Specification-anchored:} the expected value is composed in the test from
constants, published procedures or specification values fixed outside the mutate
target.\\
\textbf{State-anchored:} the expected value is obtained, directly or transitively,
from the running system under test.
\end{quote}

Of the four channels, only the expected value is named in the prior literature.
The width of a tolerance band, a conditioning variable read at evaluation time and
a scenario-placing generator are not, and the conditioning variable turns out to
carry most of the cost.

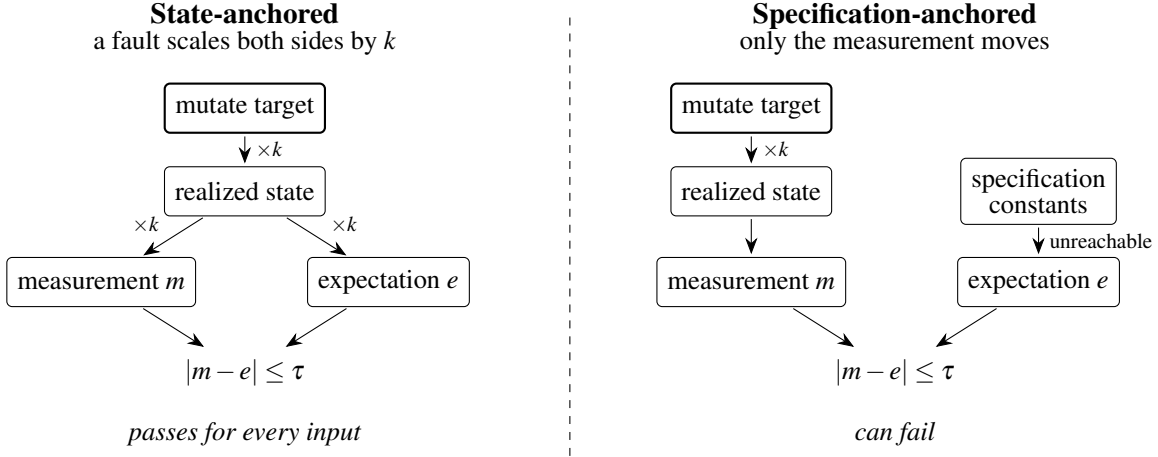
\begin{figure}[t]
\centering
\begin{tikzpicture}[
  font=\footnotesize,
  box/.style={draw, rounded corners=2pt, minimum height=6.5mm, inner xsep=4pt, align=center},
  mut/.style={box, thick},
  flow/.style={-{Stealth[length=2.2mm]}, shorten >=1pt, shorten <=1pt},
  lbl/.style={font=\scriptsize, inner sep=2pt},
]
\node[align=center] at (0,4.05) {\small\textbf{State-anchored}\\[-2pt]a fault scales both sides by $k$};
\node[mut] (mA)    at ( 0  , 3.0) {mutate target};
\node[box] (sA)    at ( 0  , 1.9) {realized state};
\node[box] (measA) at (-1.9, 0.7) {measurement $m$};
\node[box] (expA)  at ( 1.9, 0.7) {expectation $e$};
\node      (cmpA)  at ( 0  ,-0.5) {$|m-e| \le \tau$};
\node      at ( 0  ,-1.3) {\itshape passes for every input};
\draw[flow] (mA) -- node[lbl, right=2pt] {$\times k$} (sA);
\draw[flow] (sA) -- node[lbl, pos=0.6, above left=1pt] {$\times k$} (measA);
\draw[flow] (sA) -- node[lbl, pos=0.6, above right=1pt] {$\times k$} (expA);
\draw[flow] (measA) -- (cmpA);
\draw[flow] (expA)  -- (cmpA);

\draw[dashed] (4.3,-1.6) -- (4.3,4.3);

\node[align=center] at (8.6,4.05) {\small\textbf{Specification-anchored}\\[-2pt]only the measurement moves};
\node[mut] (mB)    at ( 6.7, 3.0) {mutate target};
\node[box] (sB)    at ( 6.7, 1.9) {realized state};
\node[box] (kB)    at (10.5, 1.9) {specification\\[-2pt]constants};
\node[box] (measB) at ( 6.7, 0.7) {measurement $m$};
\node[box] (expB)  at (10.5, 0.7) {expectation $e$};
\node      (cmpB)  at ( 8.6,-0.5) {$|m-e| \le \tau$};
\node      at ( 8.6,-1.3) {\itshape can fail};
\draw[flow] (mB) -- node[lbl, right=2pt] {$\times k$} (sB);
\draw[flow] (sB) -- (measB);
\draw[flow] (kB) -- node[lbl, right=2pt] {unreachable} (expB);
\draw[flow] (measB) -- (cmpB);
\draw[flow] (expB)  -- (cmpB);
\end{tikzpicture}
\caption{Why a state-anchored oracle cannot fail. On the left the expectation is
obtained from state the mutate target produces, so a fault scaling that state by
$k$ scales measurement and expectation alike and the comparison is invariant. On the
right the expectation is composed from values the target cannot reach, so the same
fault moves one side only.}
\label{fig:mechanism}
\end{figure}

\paragraph{The mechanism, stated as a cancellation.} Figure~\ref{fig:mechanism}
gives the shape and the algebra follows. Let an oracle assert
$|m - e| \le \tau$, or equivalently $m/e \in [1-\epsilon, 1+\epsilon]$, where $m$
is a measurement taken from a run and $e$ is the expectation. Suppose $e$ is
obtained by calling the module under mutation, or is a function of a state
variable that module produces, and suppose a mutation scales the underlying
quantity by a factor $k$. If the measurement is also proportional to that
quantity, then the mutant run yields $m' = km$ and $e' = ke$, so
$m'/e' = m/e$ exactly. The comparison is invariant under the mutation. The
property does not merely fail to fail on the inputs that were generated; it
cannot fail on any input, because the defect lies in the oracle rather than in
the input space. No quantity of generated traces can reveal it, and this is why
the suite in question had run green on every commit for a month.

This is the sense in which our second motivating defect and our measurement are
the same phenomenon seen twice. The pattern is that a property is blind to any
fault whose effect is absorbed into the state its oracle conditions on. In Voas's
terms the infection occurs but does not propagate to the observation point,
because the observation point is defined relative to state the infection has
already altered. \emph{Conditioning an oracle on realized state is much of what
makes an emergent-behavior property readable and stable, and it is also what can
destroy its sensitivity to faults upstream of that state.}

\paragraph{The sizing condition, which is what makes this predictive.} A claim
that self-reference costs detection would be false 3 times out of 6 in our
own data, and we report the measurement that establishes that in
Section~\ref{sec:measuring}. The narrower claim the data support is:

\begin{quote}
Self-reference costs detection \textbf{when the anchored value sizes the
comparison the oracle makes and the fault moves it proportionally}, so that
measured and expected cancel. Where the anchored value only places a scenario,
feeds a bound carrying large slack, or measures a quantity another suite covers
directly, the self-reference is \emph{latent}: real, statically identifiable, and
free.
\end{quote}

The debounce negative control sharpens this further, and the refinement was
predicted before it was measured. In that suite the comparison-operator mutants
keep dying even after the oracle is state-anchored, because the test restates
that comparison against a constant the mutation cannot reach. Faults that move
the anchored quantity cancel; faults that change a comparison the test
independently restates do not.

\paragraph{Model-freedom is the risk condition.} The strongest property-testing
results on stateful behavior in the literature are model-based
\cite{hughes}, and an abstract model is an independent source of expected values,
so that style is structurally incapable of the defect. We were explicitly
model-free, and model-freedom is precisely the absence of an independent source
of expected values, so the author reaches for the running system. The trap was
not that specification anchoring was unavailable. Air traffic control publishes
its procedural specification, and several of the values we needed already sat in
the configuration module as named constants. State anchoring was simply the more
natural thing to write, and Section~\ref{sec:measuring} measures how often it was
written.

\paragraph{The scope condition.} State anchoring cancels only when the source of
the observed state lies inside the mutate target. Reading realized positions and
speeds from a simulator to judge an autopilot is safe, because the plant and the
system under test are separate artifacts \cite{li2024}. In our simulator they
coincide by construction: the integrator that produces the trajectory is the code
being mutated. This is a scope condition rather than a caveat, and stating it is
what makes the claim falsifiable in the right place.

\section{Measuring the mechanism}
\label{sec:measuring}

\subsection{Re-anchoring recovers 8 mutants}

The holding suite's two containment properties assert that the steady-state
racetrack width stays within a tolerance band of twice the aircraft's turn
radius. The band is two-sided on purpose. A pattern wider than the band leaves
the protected airspace the hold is meant to stay inside, a pattern narrower than
it never opens out to that airspace at all, and both must fail. That form is
required by our own methodology. The expected width, however, was computed by calling the
turn-radius function exported by the module under test, evaluated at the
aircraft's realized true airspeed.
Figure~\ref{fig:hold-geometry} drew the consequence to scale, and is the one piece
of domain background the rest of this section assumes; it is worth turning back to.

\begin{figure}[t]
\begin{tcblisting}{codebox}
// BEFORE. Both the yardstick and its speed come from the code under test.
import { holdTurnRadiusNM, holdOutboundLegSec } from '@/core/holdGeometry.ts';

twoR: 2 * holdTurnRadiusNM(ac.trueAirspeedKt),   // <- realized state
legTimeSec: holdOutboundLegSec(ac.position.z),

// H4, unchanged by the repair:
expect(maxXte).toBeLessThanOrEqual(1.06 * run.twoR);
expect(maxXte).toBeGreaterThanOrEqual(0.94 * run.twoR);
\end{tcblisting}
\begin{tcblisting}{codebox}
// AFTER. The yardstick is composed in the test from the published constants,
// and the speed is the specification's speed for the altitude, not the
// aircraft's realized speed. No assertion is added and no source is changed.
import { specTurnRadiusNM, specHoldTasKt, specOutboundLegSec }
  from './holdSpecGeometry.ts';

twoR: 2 * specTurnRadiusNM(specHoldTasKt(s.iasKt, s.altFt)),
legTimeSec: specOutboundLegSec(ac.position.z),

// holdSpecGeometry.ts, which deliberately does NOT import holdGeometry.ts:
export function specTurnRadiusNM(tasKts: number): number {
  const bankRad = (SIMULATION.HOLD_MAX_BANK_DEG * Math.PI) / 180;
  const bankLimitedDegPerSec = (G_BANK_FACTOR * Math.tan(bankRad)) / tasKts;
  const rateDegPerSec = Math.min(
    SIMULATION.HOLD_TURN_RATE_DEGS_PER_SEC, bankLimitedDegPerSec);
  return (tasKts * KNOTS_TO_NM_PER_SEC) / ((rateDegPerSec * Math.PI) / 180);
}
\end{tcblisting}
\caption{The containment yardstick before and after re-anchoring. The repair
changes where the expected value comes from and nothing else: the same two
assertions, the same tolerance band, the same generator, the same pinned seed,
the same 46 mutants. It recovers 9 mutants, 8 of them from the yardstick alone.}
\label{fig:hold}
\end{figure}

We closed the anchoring channels one at a time against a fixed 46-mutant
population, with seeds pinned throughout so every figure is deterministic rather
than sampled. Table~\ref{tab:ablation} reports the result.

\begin{table}[t]
\centering\small
\begin{tblr}{colspec={llrrr}, row{1}={font=\bfseries}}
\toprule
Run & Channels closed & Killed & Score & P-only \\
\midrule
A0 & as found                                        & 12 & 26.09\% & 0 \\
A1 & expected value recomposed                       & 13 & 28.26\% & 0 \\
A2 & plus generator recomposed                       & 13 & 28.26\% & 0 \\
A3a & plus yardstick re-anchored on specification    & 21 & 45.65\% & 1 \\
A3b & A1 plus a new speed-bound oracle               & 21 & 45.65\% & 1 \\
A5 & every closure plus the new oracle               & 21 & 45.65\% & 1 \\
\bottomrule
\end{tblr}
\caption{Ablation over a fixed population of 46 mutants on the holding-geometry
module. A3a adds no assertion; it changes only where the expected value comes
from. A3b changes no existing oracle; it adds one. The two recover the identical
8 mutants and applying both together adds nothing.}
\label{tab:ablation}
\end{table}

\paragraph{Two recovery counts, and which is which.} Two counts run through the
rest of this paper and they are not interchangeable. Closing \emph{every}
anchoring channel on this module recovers \textbf{9} mutants, taking A0's 12 to
A3a's 21, and that is the doubling of the score from 26.09 to 45.65\%. Closing
the \emph{single} conditioning instance recovers \textbf{8} on its own, taking
A1's 13 to A3a's 21.

Three things in that table matter more than the headline score.

\begin{enumerate}
\item \textbf{A3a adds no assertion.} It removes exactly one value flow, the
realized true airspeed feeding the containment yardstick, and recovers 8 mutants
on its own, so the blind spot was caused by the conditioning and not by a missing
test. This is the provenance the test-smell rule legislates over
\cite{rwemalika2021}, moved and nothing else. Figure~\ref{fig:hold-geometry} draws why: anchored on realized speed, a
speed-table mutation moves the expected racetrack width along with the flown one;
anchored on the specification's speed for the altitude it does not.

\item \textbf{The two remedies are redundant rather than additive.} That is what
one expects if a single mechanism is responsible, and it is not what one expects if
the suite simply had two unrelated gaps.

\item \textbf{The first reading of this ablation concluded the opposite.} An
earlier analysis argued that the third channel was not a self-reference closure
at all, on the grounds that the new oracle adds an assertion rather than removing
a value flow, and concluded that self-reference was nearly costless at one
mutant. The A3a run refutes that, and we report the sequence because it is the
same species of error the rest of the paper is about: a plausible argument that
was not checked against a measurement it could have been checked against.
\end{enumerate}

\paragraph{The 1-against-8 framing is not the right comparison, and we do not make it.}
It is tempting to read the ablation as saying that the channel the literature has
named costs one mutant while the channel it has never named costs 8. That
ratio is partly an artifact of mutant density. The 8 occupy 3 source
lines, and 6 of them sit on a single nested ternary expressing one symptom, a
wrong commanded hold speed; the ternary admits 4 \texttt{ConditionalExpression}
and 4 \texttt{EqualityOperator} mutants where the first channel's site, a
\texttt{Math.min} call, admits exactly one replacement.

We ran the subsumption analysis rather than estimating it, and the answer is worse
than the estimate we had made. Re-running the population with bail disabled yields
a full kill matrix, and all 8 recovered mutants share an identical killer set, the
same 3 tests, so under dynamic subsumption they form a single class. They are one
symptom counted eight ways. The negative control of
Section~\ref{sec:negative} behaves the same way: the 4 mutants state anchoring
costs fall into 2 classes, of which one subsumes the other. Two caveats travel with
that. The relation is dynamic and therefore relative to the suite that computed it,
and over 13 tests it is coarse, so two mutants share a killer set because this
suite cannot separate them rather than because they are one fault; the survey says
plainly that the true disjoint set is not computable and only an approximation is
available \cite{papadakis2019}. And the analysis needs bail disabled, which the
runs behind every other figure in this paper did not use, so it is a separate
measurement over the same two populations rather than a re-reading of the
published ones.

The comparison that does not depend on how many mutants a line admits is the
property-only column. Closing the named channel moved it by 0, because the example
suite already killed that mutant. Closing the unnamed one moved it from 0 to
1, and that mutant is the 14{,}000 foot band boundary, the only property-only mutant
anywhere on this module. 0 against 1 is a weaker-sounding claim than 1
against 8, and it is the one the measurement supports.

\subsection{State-anchoring a healthy oracle costs 4 mutants}
\label{sec:negative}

The removal alone is a repair, and a repair is consistent with several
explanations. The converse experiment is what makes the mechanism an intervention.
The debounce property suite is the reference specification-anchored case at 18 of 19,
its assertions comparing against \texttt{ELIGIBILITY\_HOLD\_SEC}, imported from
outside the mutated range and therefore unperturbable by any mutant in the
population.

We deliberately state-anchored it, replacing the streak the test recomputes from
its own input trace with the controller's own accumulated timer read from the
object under test. This is a design a reviewer would wave through, since
recomputing in the test what the object already tracks looks redundant, and that
plausibility is the point. No production source was changed and no assertion was
removed.

\begin{figure}[t]
\begin{tcblisting}{codebox}
// BEFORE (specification-anchored). The test accumulates the streak itself from
// the trace it generated, and compares against an imported constant.
let ineligibleStreak = 0;
for (const step of steps) { /* ... */ ineligibleStreak += step.dt; /* ... */ }
expect(ineligibleStreak).toBeLessThan(HOLD_SEC);          // HOLD_SEC from :37

// AFTER (state-anchored). The oracle reads the accumulator the controller
// maintains, which is inside the mutate range. A missing entry reads as
// Infinity, because a released hold correctly deletes its map entry.
const held = controller.eligibilityHold.get(ac.id);
expect(held?.ineligibleSec ?? Infinity).toBeLessThan(HOLD_SEC);
\end{tcblisting}
\caption{The debounce negative control. Detection falls because the oracle now
reads the value the mutation perturbs. The direction was predicted in advance;
which mutants would survive was predicted only in part, as the text sets out.}
\label{fig:debounce}
\end{figure}

Detection fell from 18 of 19 to 14 of 19, from 94.74 to 73.68\%. Because
two records of this prediction exist and they differ, its provenance is worth
stating precisely.

Two predictions exist and they differ. The first was written into the measurement
plan before the state-anchored oracle was written, and named lines 112 and 113,
the accumulator and the expiry comparison. The second was recorded after the
oracle was written but before it was run, and refined the first by predicting that
the comparison-operator mutants at 113 would keep dying, because the test's own
comparison is against a constant outside the mutate range that no mutation can
move.

Against the outcome, the record is this. Four mutants stopped dying: the
accumulator at 112, one branch-forcing mutant at 113, and two at 108 in the
external-clear guard. The accumulator was predicted. The refinement about the
comparison operators was correct and is the sharpest available statement of the
sizing condition, since it distinguishes faults that move the anchored quantity
from faults that change a comparison the test independently restates. But the
first prediction was wrong about 113 in its unrefined form, and the pair at 108,
half the effect, appears in neither prediction. We therefore claim that the
direction and magnitude were predicted and that 2 of the 4 surviving mutants
were named, not that the prediction held mutant for mutant. The state-anchored
suite was then discarded; the working tree keeps the specification-anchored
version.

Removing state anchoring recovers 8 mutants on holding geometry; introducing it
costs 4 on the debounce.

\subsection{A reference model kills what specification anchoring kills}

Our argument that model-freedom is the risk condition rested on a structural
claim plus one external citation, which is thin for a claim that carries this much
of the paper. The model-conformance file of Section~\ref{sec:system}, the one the
\texttt{*.property.test.ts} rule excludes, is the within-system control we had not
run: it maintains a reference model of the debounce state machine, asserts that the
real controller matches it step for step over the same 19-mutant population, and is
the design our argument holds to be structurally incapable of the anchoring
defect.

Run against that population it kills 18 of 19, and the diff against the
specification-anchored invariant suite is empty in both directions: the same 18
mutants die, the same one survives, and that survivor is the guard we argue is
equivalent. Table~\ref{tab:control} reads the three designs over that one fixed
population:

\begin{table}[h]
\centering\small
\begin{tblr}{colspec={llrr}, row{1}={font=\bfseries}}
\toprule
Design & Expectation comes from & Killed & Score \\
\midrule
Reference-model conformance   & an independent model      & 18 / 19 & 94.74\% \\
Specification-anchored properties & imported constants    & 18 / 19 & 94.74\% \\
State-anchored properties     & the object under test     & 14 / 19 & 73.68\% \\
\bottomrule
\end{tblr}
\caption{The same 19 mutants judged by 3 oracle designs. The model-based and
specification-anchored designs are indistinguishable, killing the identical set;
only state anchoring loses mutants.}
\label{tab:control}
\end{table}

This sharpens the claim. A reference model does
not detect more here than a property suite with no model at all, provided the
property suite anchors its expectations outside the mutate target. Read against
the oracle-assessment literature, which scores an oracle by the faults it fails to
reveal \cite{jahangirova2016,jahangirova2018tse}, the two designs are
indistinguishable on this population. What separates
the designs is not whether a model is present but where the expectation comes
from, which is the paper's thesis measured on one module instead of argued from
the literature. It also narrows our own advice: the reason to prefer a model is
maintainability or expressiveness, not fault detection, and the reason to fear
model-freedom is that it removes the obvious independent source of expectations
and leaves the running system as the nearest substitute.

We report two caveats. This is one module and the smallest population in the
study, and the model suite achieves its result with 2 tests against the invariant
suite's 4, so the comparison is of designs rather than of effort.

\subsection{Prevalence, and why the cost figure must travel with it}

A static pass over all 12 invariant suites, asking whether each obtains a
value from a module under mutation and uses it as an expected value, as a
conditioning variable read at evaluation time, or as a generator, found 6
instances in 3 suites. Two were previously unknown. All 6 have now been
ablated against fixed populations, and the result is the reason the sizing
condition exists.

\begin{table}[t]
\centering\small
\begin{tblr}{colspec={llr}, row{1}={font=\bfseries}}
\toprule
Instance & Channel & Recovery \\
\midrule
Yardstick from the turn-radius function        & expected value              & +1 \\
Generator from the turn-radius function        & scenario placement          & +0 \\
Yardstick from realized true airspeed          & conditioning, sizes the band & \textbf{+8} \\
Leg time from the outbound-leg function        & expected value, bound with slack & +0 \\
Angle measurement from the shared math module  & measurement, covered elsewhere & +0 \\
MSAW alert against realized vertical speed     & conditioning, is the expectation & \textbf{+3} \\
\bottomrule
\end{tblr}
\caption{Six state-anchored instances ablated against fixed mutant populations.
The two that satisfy the sizing condition carry 11 of the 12 recovered
mutants; the other 4 carry one between them. The prevalence figure should not be
reported without this split. The last row's recovery is conditional on the mutate scope and is the
subject of Section~\ref{sec:scope}.}
\label{tab:prevalence}
\end{table}

Six instances found in 3 of 12 suites means the defect is not
idiosyncratic, and that the claim about state anchoring being the more natural thing
to write is measured rather than argued. 2 of 6 carrying almost the whole cost
means the prevalence figure would badly overstate the mechanism if it travelled
alone; the honest summary is 6 found, 6 measured, 2 costly. The split is not
arbitrary: the two costly instances are exactly the two where the anchored value
sizes the comparison the oracle makes, and they carry 11 of the 12 mutants
the ablations recover. The count of 6 is still a floor, because channels reached
through an import are found reliably while judging whether a value is realized state
required reading each oracle, which is judgment rather than analysis.

\subsection{Scope relativity, measured}
\label{sec:scope}

The sixth instance in Table~\ref{tab:prevalence} is the one whose anchoring is
decided by the mutate scope rather than by the oracle text. The paper has so far
asserted that scope relativity on one citation \cite{li2024}. This measures it.

The MSAW suite asserts that a terrain alert reflects the aircraft's current
vertical direction, against a terrain service wired to be unsafe exactly while
descending:

\begin{tcblisting}{codebox}
expect(ac.msawAlert).toBe(ac.verticalSpeed < 0);
\end{tcblisting}

Both sides read \texttt{ac.verticalSpeed}. The terrain service returns
\texttt{currentSafe = predictedSafe = (vs >= 0)} and the physics block turns that
into the alert; the expectation is the same variable read at evaluation time. The
anchored value is not merely one input to the comparison, it \emph{is} the
expected value, so the sizing condition is satisfied as sharply as it can be.

What makes the instance decisive is that \texttt{ac.verticalSpeed} is produced by
\texttt{updateAltitude} at lines 856 to 925, while the MSAW block the suite exists
to test is lines 768 to 799. Same file, different regions. So we ran the identical
oracle against two declared scopes, changing nothing else:

\begin{table}[h]
\centering\small
\begin{tblr}{colspec={llrrrr}, row{1}={font=\bfseries}}
\toprule
Cell & Mutate scope & Anchor in scope & State-anchored & Spec-anchored & Recovery \\
\midrule
S1 & 768--799            & no  & 21 / 57 (36.84\%)  & 21 / 57 (36.84\%)  & 0 \\
S2 & 768--799, 856--925  & yes & 39 / 121 (32.23\%) & 42 / 121 (34.71\%) & \textbf{+3} \\
\bottomrule
\end{tblr}
\caption{One oracle, two declared mutate scopes. Re-anchoring is inert when the
value's producer lies outside the target and recovers 3 mutants when it lies
inside. Scores across cells are not comparable, since S2's population contains
S1's; only the within-cell recovery is.}
\label{tab:scope}
\end{table}

In S1 the diff is empty in both directions: the same 21 mutants die. In S2 3
mutants are recovered, and the location of all 3 is
the result. Not one is in the MSAW block. They are a branch inversion at 866 that
sends a commanded climb down the descent path, and two rate computations at 910
and 911 that drive the aircraft to its target inside the trace so that it levels
off and the vertical speed goes to zero. In each case the state-anchored oracle
recomputes its expectation from the vertical speed the mutation has just moved,
and agrees with itself. The MSAW block's own 21 killed and 28 survived are
identical under both anchorings.

\emph{Re-anchoring bought detection only against faults in the code that produces
the anchored value, and none against faults in the code the oracle was written to
test.} That measures the scope condition instead of merely stating it: an oracle
is not state-anchored in the abstract, it is state-anchored with respect to a
declared target.

We recorded 4 predictions before writing the re-anchored oracle. The first, that
S1 recovery would be zero, held more strongly than stated, since the diff is empty
rather than merely equal in count. The second, that S2 recovery would be positive,
held in direction and magnitude and was wrong about location: we named the
vertical-speed assignments, and the mutants that moved were the branch selector and
the two rate computations feeding them. The third, that anchoring would not disturb
the MSAW block, held exactly. We report the missed addresses because the mechanism
and the addresses are separate claims and only one was predicted.

\subsection{The published rule would revert the repair}

Our repair composes the expected value in the test from specification constants,
in a helper that deliberately does not import the module under mutation and that duplicates the regulatory speed literals instead of importing the function that serves them. That helper is emphatically a value \emph{computed during the test},
which is precisely what the published detector forbids: the expected value should
be a constant or a reference to a constant and not computed during the test
\cite{rwemalika2021}, and the catalogue entries \texttt{On The Fly} and
\texttt{Second Guess The Calculation} say the same \cite{soares2023catalog}.
Applying the published smell rule to our repaired suite would flag the entire
helper and revert the fix that recovered 8 of 9 mutants.

The honest framing is not that the published rule is wrong. It uses "computed in
the test" as a proxy for "duplicates production logic", which is a real
maintainability cost that our helper does incur: if the specification changes,
two places change. The rule optimizes maintainability and, taken literally,
degrades fault detection. Nobody had measured that exchange rate. On this module
it is 9 mutants of 46.

This is also why we state the predicate as provenance rather than as computedness.
What determines killability is not whether the expected value was computed during
the test but whether it flows from the module under mutation, so a value composed in
the test from constants fixed outside the mutate target is unperturbable and safe,
which is exactly the case the published rule rejects.

\section{Writing the missing oracle, and what it caught}
\label{sec:prospective}

Everything above is retrospective. Section~\ref{sec:results} reported that
collision avoidance is close to untested by both suites, that its survivors
concentrate where the advisory's content is decided, and that coordinated
opposite senses, the defining property of the concept, is constrained by no
oracle in either suite. It also said this is the one module where a competent
second author would do materially better. That is a prediction about where a
defect would be found if anyone looked, and it can be tested, so we looked.

\paragraph{The oracle found a live defect.} A property was added asserting that
the two aircraft of a coordinated advisory receive opposite senses and that each
is commanded in the sense it was given. It failed against the deployed code.
\texttt{resolveDirections} does return an opposite pair in every branch, but a
later step overrode the sense to \emph{climb} for any aircraft on final approach,
one aircraft at a time and with no knowledge of its partner. A pair that were
both on approach were therefore both commanded to climb, and since each climbed
by the same fixed offset the advisory preserved the vertical gap it had fired on
exactly. The override was unreachable in every other case, because the same
aircraft already outranked its partner in sense selection; it changed the outcome
only in the case where it destroyed coordination.

Two geometries reach it. Two aircraft compressing on one final need an unrealistic
overtake, but two aircraft on final to \emph{different} airports 1.4~NM and 300~ft
apart is ordinary terminal geometry, and the parallel-approach exemption does not
cover it because that exemption is keyed on the airport.

\paragraph{The instrumentation could not have seen it.} The event stream and the
log both reported the senses \texttt{resolveDirections} returned, which are opposite by construction, in place of the senses applied. Across 147 recorded
activations the telemetry showed complementary senses 100\% of the
time, and it would have shown exactly that in a world where every advisory was
same-sense. This is the same defect as the paper's subject, one layer further
out: an observation derived from the same source as the expectation cannot
contradict it. We report it because it bears directly on prevalence. A reader may
reasonably ask why a defect of this size was never noticed in operation, and the
answer is that the only instrument pointed at it was state-anchored.

\paragraph{A second defect, from reading the standard.} Writing an oracle for
\emph{which} aircraft receives which sense required the selection rule rather
than our own logic. Airborne collision avoidance system (\textsc{acas}) guidance selects the sense
giving the greater
vertical distance and, where an altitude crossing is projected, the sense that
avoids crossing. Our tiebreak did the opposite, sending the lower aircraft up
through the higher one. Because the tie is the common case, crossing advisories
were the default behaviour.

\paragraph{Two rounds, and the one that failed is the informative one.} Both
rounds recorded predictions before the runs, in the same committed form as
Section~\ref{sec:measuring}. The first predicted a score of 32 to 48\% and scored
1 of 5: the measured figure was 29.81\%, and \texttt{directionScore}, which we had
named as the function that would move, recovered exactly zero mutants. The
diagnosis was that the new property constrained the \emph{relation} between the
two senses and not their \emph{assignment} to aircraft, so a mutation that swaps
which aircraft climbs still satisfies it. That is not anchoring. It is the
adjacent failure Section~\ref{sec:system} warns about, a one-directional oracle
satisfied by code that does too little, and the two should not be conflated.

The second round tested that diagnosis and scored 6 of 6, including its stated
falsifier: \texttt{directionScore} moved for the first time, by 8 of its 23. The
sharpest number is not the score but a control. Inverting the sense assignment
reverses which aircraft climbs in every tied encounter, and the pre-existing suite,
3{,}824 tests by the time of this round, passed unchanged, while the 5-property suite scored
\emph{identically, to the mutant}, before and after. A semantic reversal of the
module's primary output was invisible to everything that existed to check it.

\paragraph{What this does and does not establish.} It does not establish
prevalence. It is one module of one system, and the module was chosen by this study's own survivor clustering, not at random. That selection is the result's mechanism: the analysis said where to look and
a defect was there. What it establishes is that the clustering is diagnostic
rather than descriptive. Both defects reached deployment, neither was found by
use, and both were found by writing the oracle the analysis said was absent.

\section{Threats to validity}
\label{sec:threats}

\subsection{Internal validity}

\paragraph{Scores are not random variables here.} A score over a
generator-driven suite is in principle a random variable, and we carried that as a
threat until we measured it. Ten repetitions across the two suites that pin no
seed returned identical kill counts, scores and property-only sets, down to which
mutants were property-only. At 150 to 400 cases per property detection is close to
binary: a mutant that changes anything an oracle examines dies on essentially
every draw and one that does not dies on none, so no borderline population remains for a seed to move. Ravi and Coblenz found 96\% of their mutations within
350 inputs \cite{ravi2025}. This is evidence, not proof, covering two suites over
two modules plus a 3-run check on a third, and it should not be confused with
a historical flake in this project's bearing-reciprocity property: that was a
defect \emph{in the oracle} surfacing on rare inputs, not a mutant escaping
detection. Seed sensitivity exists here for oracle defects while mutant detection
is deterministic.

\paragraph{The partition, and a conservatism claim that does not survive.} A test
partition assigned by reading imports invites the argument that sweeping
incidental files into the example category makes the study conservative with
respect to its claims about properties. We made that argument to ourselves and it
does not hold: the rule it assumed was applied to two modules and inverted on the
other two, so the direction of the bias was not consistent enough to be claimed in
either direction. The coverage-derived partition of Section~4 replaces it and is
reproducible from the reports themselves rather than from a reading of the test
files.

\paragraph{Mutant counts are not independent evidence.} Discarding subsumed
mutants is named as best practice for empirical mutation studies
\cite{papadakis2019}. We ran it on the two populations that carry the
interventions, and it confirms the concern rather than relieving it. The 8 mutants
re-anchoring recovers form one subsumption class and the 4 state anchoring costs
form two, so a count of recovered mutants substantially reflects how many
replacements a syntactic site admits. Section~\ref{sec:measuring} reports both and
gives the density-independent comparison alongside them. We did not run it on the
other two populations, where no intervention is claimed. The 3 zero-recovery instances
in Table~\ref{tab:prevalence} are unaffected, since zero is zero under any
counting rule.

\paragraph{The operator set cannot reach specification constants.} Tool choice
is itself a documented source of divergence between mutation studies
\cite{papadakis2019}. Ours is StrykerJS throughout, and it
ships no numeric-literal mutator, so the configuration constants our repaired
oracles compose from cannot be perturbed by any mutant in these populations. Our "specification-anchored" therefore overlaps with "anchored on something this
tool cannot mutate." We think the overlap benign, since mutating a specification constant mutates the specification and not the implementation and an oracle that tracks it is behaving correctly, but we
did not measure that and a tool with a literal mutator would be needed to.

\subsection{Construct validity}

\paragraph{Equivalent mutants and comparability.} Our equivalence treatment is
described in Section~4 and is exhaustive on only two of 4 modules. Our
denominator includes uncovered mutants and our timeouts count as killed, both
Stryker defaults; under the covered-only convention \cite{schuler2011} the as-found holding property figure reads 41.38 in place of 26.09 and the spread
across the 4 modules narrows from 4.63 to 3.49. We publish both conventions and defend neither. No retrieved study keys mutant identity as we do, so the
survivor-diff columns cannot be validated against anyone else's, and any
cross-paper comparison of our absolute percentages is meaningless without these
statements.

\paragraph{Mutation score is a proxy, and a contested one.} Papadakis et al.\ show over a large corpus that correlations between mutation scores and
real fault detection are weak once test suite size is controlled for
\cite{papadakis2018}, and their survey makes controlling for it standing advice
in empirical mutation studies \cite{papadakis2019}. That finding bears on our two kinds of result very
differently, and the distinction is the reason we report them separately. The
mechanism results are ablations over a fixed mutant population with suite size
held constant: the re-anchoring adds no assertion and removes none, the negative
control changes an expectation's provenance and nothing else, and the
reference-model comparison holds the population fixed while varying only the
oracle design. A difference in suite size cannot produce a difference between two runs that have
the same tests. The property-only result is exposed to it, because
the two suites differ by more than an order of magnitude in test count, 4 against
81 on the debounce and 43 against 1{,}708 on the angle module. That is why we
report kills per test alongside it, and why the property-only count should be read as a statement about these two suites and not about the two styles.

\subsection{External validity}

\paragraph{Single system, single author.} All measurements come from one
simulator whose code, invariant suites and example tests were written by the same
person, who also designed this study. A claim about the relative effectiveness of
two testing styles is exactly the kind of claim one author's habits can
determine, and we have no defense that removes this. Three amendments are owed.
The first widens it: that person worked throughout with a language model assistant
(Section~\ref{sec:design}), and on the holding-geometry module carrying this
paper's central result the production code and both covering suites were authored
in sessions with the same model, so oracle and code under test share an authoring
source. The second cuts the other way, since a coding agent wrote the ablation
repairs to a specification fixed in advance, which changes the authorship of the
measured artifact for those runs. And this is
a simulator, not certified avionics; the honest description is a deployed system
with emergent temporal behavior and an external procedural specification, and we
do not reach for safety-critical framing. These threats fall hardest on the property-only result and least on the mechanism, because a cancellation
argument and a controlled ablation over a fixed population do not depend on who
wrote the code.

\paragraph{Four modules.} Four is few, and they were chosen to span the
structural range rather than sampled. We deliberately did not add more. The
mechanism claim is algebraic and does not improve with more samples, and the binding constraint on the property-only result is one author and one
codebase, not the module count; more modules would move it from 3 mutants of
366 in one system to some larger number in the same one system. The ablations are
a separate matter and there we did add a module, since
Section~\ref{sec:scope} needed a fifth mutate target to vary the scope against.

\section{Propagation distance as a rival explanation}
\label{sec:distance}

Anchoring is not the first explanation we reached for, and it is not the first
one a reader will reach for either. The obvious rival is propagation distance:
the further a fault must travel from the mutated code to the point where an
oracle looks, the more chances it has to be absorbed, so oracles far from the
code they judge should detect less. That is a real effect, it is the standard
reading of Voas's model \cite{voas1992}, and it was our working explanation for
most of this study. We set it out here with the reasons we abandoned it, because
a reader who suspects the two hypotheses are the same relabelled is owed the comparison.

The first reason is that we never computed it. Our ordering of the modules was
assigned by reading the code and describing some oracles as nearer to the fault
than others. No distance was measured, and an explanation that is asserted by
narration explains nothing.

The second is that the quantity is not open, so measuring it would have been replication and not explanation. Niedermayr and Wagner define the minimal
stack distance between test case and method and correlate it with mutation-based
effectiveness across 21 projects \cite{niedermayr2019}; Vera-P\'erez et al.\ compute the same measure and relate it to their symptom diagnosis
\cite{veraperez2019}; and Jahangirova et al.\ use the term itself
\cite{jahangirova2016}.

The third is that the contrast which made distance look explanatory was an
artifact of the partition. The ratio we had in hand was 94.74 over 28.26, a
factor of 3.4 between two named modules, and it rested on a debounce example
score of 21.05\% that the coverage-derived partition of Section~4 raises to
89.47. The corrected spread across all 4 property scores is 4.63, and the pair
ratio was in any case never the range.

The fourth reason is the one that settles it, and it is a measurement and not an argument. Section~\ref{sec:scope} runs one unchanged oracle against two mutate
scopes. The oracle, the harness, the aircraft and the code producing the value it
reads are identical in both cells and stand in the same call relation, so whatever
propagation distance means, it does not differ between them. Distance therefore
predicts the same outcome in both cells. Anchoring predicts opposite outcomes,
and the opposite outcomes are what we observed. \emph{The two hypotheses are
distinguishable and the data pick one.}

What survives from the distance intuition is that it was reaching for something real
with the wrong variable. The ordering it described is genuine; what determines that
ordering is not how far the fault must travel but whether the yardstick travels with
it, and anchoring predicts that ordering while also predicting the three
interventions in Section~\ref{sec:measuring}, which distance does not.

Neither hypothesis is the whole story, and Section~\ref{sec:prospective} put a
number on a third mechanism. Proving the new oracles could fail required, twice,
disabling two sites rather than one: the ground-proximity inhibit tests each
aircraft in a separate clause, and the parallel-approach exemption is implemented
in both the conflict detector and the advisory controller. A single-site mutation
in either pair survives an oracle that would otherwise catch it, because the
redundant copy still holds. Redundant guards are a mechanism for survival that is
neither anchoring nor distance, it is invisible to both predicates, and any
prevalence figure that attributes all survivors to a single cause will overcount.
We did not measure how much of our own population it accounts for, and a study
that separated the three would be worth more than another that argued between
two of them.

\section{Implications}
\label{sec:implications}

The recommendation we draw is narrower than an endorsement of property-based
testing and narrower than a criticism of it.

\paragraph{Check the anchoring before writing more properties.} It is a static
question, it takes minutes per suite, and it does not require a mutant
population. For each assertion, ask where the value on the expected side came
from, and treat a value that flows from the module under mutation as a defect
unless the anchored quantity does not size the comparison. The flow half of that
predicate is decidable on the import-and-call graph, and no shipping linter or
test-smell detector we swept implements even that much, which is a tractable tooling gap and not a research problem. The sizing half is not decidable and we do not
claim it is: it took us a reading of each oracle.

Two qualifications belong with that advice. The first is that a practitioner has
no mutate target. Our predicate is defined relative to a declared mutate scope,
and a reader not running mutation testing has none. What it reduces to is the
question of whether the expected value is produced by the code the test is meant
to be judging, which is a judgment about intent and not a property of a configuration; the module or package boundary is the usual proxy and is the one we
would recommend. The second is that the scope is not a formality.
Section~\ref{sec:scope} shows one unchanged oracle that is state-anchored under one
target and not under another, so a report that an oracle is anchored is incomplete
until it says anchored with respect to what.

\paragraph{Model-freedom is not itself the risk; anchoring is.} Our control in
Section~\ref{sec:measuring} shows that avoiding a reference model cost us nothing in
detection on the module where we could measure it. The cost we did not price is
subtler: model-freedom removes the obvious independent source of expected values,
and the author then reaches for the running system, which is the move that does cost
detection. Specification anchoring is the third option and it was available to us
for free: the domain publishes its procedures, and the constants were already named
in our own configuration.

\paragraph{A green property suite is not evidence that its oracles discriminate.}
Our discipline already required a hand-picked mutation per oracle, recorded in a
comment, and that practice did not surface the defect, because the author who wrote
the tautology also chose the mutation used to test it. Generated mutants are cheap
and they are chosen by something other than the author's model of their own blind
spots.

\paragraph{Report property-only counts rather than scores.} The scores in
Table~\ref{tab:main} would support several different and mutually inconsistent
stories on their own. The survivor diff carries the information, and it must be read
against the population.

\section{Conclusion}

We measured whether the invariant oracles in a deployed air traffic control
simulator detect faults, running the property suites and the example suites
separately over identical mutant populations under a coverage-derived partition.
Across 4 modules and 366 distinct mutants the property tests covering them add
3 mutants of detection over the hand-written tests that already existed, 4
once one of those suites is re-anchored, while remaining 6 to 33 times more
efficient per test, so the finding is near-total overlap and not weakness. The
cause is not care taken in writing the oracles but where their expectations are
anchored: an oracle that obtains its expected value from the system's own realized
state cannot discriminate faults in whatever produces that state, when the anchored
value sizes the comparison, because the fault moves both sides at once. We showed
this three times, recovering 8
mutants by re-anchoring one oracle on the specification without adding an
assertion, costing 4 by deliberately state-anchoring a healthy one, and turning
the same repair from inert to worth 3 mutants by moving nothing but the
declared mutate target. On the debounce population a reference-model suite killed
exactly what specification anchoring killed, which places the risk in anchoring and not in model-freedom. The defect is already named at one of its four
channels and, in the literature we retrieved, has not been measured at any of
them; the published rule for
detecting it, applied literally to our repair, would revert it. A green property
suite is not evidence that the oracles in it discriminate, and the cheapest way to find out is to generate the mutants instead of choosing them.

\section*{Availability}

The dataset, the analysis program, and a self-contained reader that recomputes
every table in this paper from the dataset with no other input are deposited at
\url{https://doi.org/10.5281/zenodo.21940547}, which resolves to the latest
version. Its README states what the dataset carries, why line and
column positions are absent from it and not derivable from it, which measurements
fall outside its scope, and an earlier version of the dataset whose identity
scheme we broke ourselves before any of it was published. The simulator is not public: every number here can be
recomputed and every diff re-derived, but no measurement can be re-run, because
re-running one requires the simulator.

{\footnotesize
\bibliographystyle{plain}
\bibliography{references}
}

\end{document}